\documentclass[12pt,preprint]{aastex}
\usepackage{graphicx}
\usepackage{multirow}
\shorttitle{Profiles and morphology of SINGS galaxies}
\shortauthors{Mu\~{n}oz Mateos et al.}

\begin{document}

\title{Radial distribution of stars, gas and dust in SINGS galaxies: I. Surface photometry and morphology}

\author{J.C. Mu\~{n}oz-Mateos\altaffilmark{1},
A. Gil de Paz\altaffilmark{1},
J. Zamorano\altaffilmark{1},
S. Boissier\altaffilmark{2},
D.A. Dale\altaffilmark{3},
P.G. P\'{e}rez-Gonz\'{a}lez\altaffilmark{1},
J. Gallego\altaffilmark{1},
B.F. Madore\altaffilmark{4},
G. Bendo\altaffilmark{5},
A. Boselli\altaffilmark{2},
V. Buat\altaffilmark{2},
D. Calzetti\altaffilmark{6},
J. Moustakas\altaffilmark{7},
R. C. Kennicutt, Jr.\altaffilmark{8,9}
}

\altaffiltext{1}{Departamento de Astrof\'{\i}sica y CC$.$ de la Atm\'osfera, Universidad Complutense de Madrid, Avda$.$ de la Complutense, s/n, E-28040 Madrid, Spain; jcmunoz, agpaz, jaz, pgperez, jgm@astrax.fis.ucm.es}
\altaffiltext{2}{Laboratoire d'Astrophysique de Marseille, OAMP, Universit\'e Aix-Marseille \& CNRS UMR 6110, 38 rue Fr\'ed\'eric Joliot-Curie, 13388 Marseille cedex 13, France; samuel.boissier, alessandro.boselli, veronique.buat@oamp.fr}
\altaffiltext{3}{Department of Physics and Astronomy, University of Wyoming, Laramie, WY; ddale@uwyo.edu}
\altaffiltext{4}{Observatories of the Carnegie Institution of Washington, 813 Santa Barbara Street, Pasadena, CA 91101; barry@ociw.edu}
\altaffiltext{5}{Astrophysics Group, Imperial College, Blackett Laboratory, Prince Consort Road, London SW7 2AZ; g.bendo@imperial.ac.uk}
\altaffiltext{6}{Department of Astronomy, University of Massachusetts, Amherst, MA 01003; calzetti@astro.umass.edu}
\altaffiltext{7}{Department of Physics, New York University, 4 Washington Place, New York, NY 10003, USA}
\altaffiltext{8}{Institute of Astronomy, University of Cambridge, Madingley Road, Cambridge CB3 0HA, UK}
\altaffiltext{9}{Steward Observatory, University of Arizona, Tucson, AZ 85721}

\begin{abstract}
We present ultraviolet through far-infrared surface brightness
profiles for the 75 galaxies in the {\it Spitzer} Infrared Nearby
Galaxies Survey (SINGS). The imagery used to measure the profiles
includes GALEX UV data, optical images from KPNO, CTIO and SDSS,
near-IR data from 2MASS, and mid- and far-infrared images from {\it
Spitzer}. Along with the radial profiles, we also provide
multi-wavelength asymptotic magnitudes and several non-parametric
indicators of galaxy morphology: the concentration index ($C_{42}$),
the asymmetry ($A$), the Gini coefficient ($G$) and the normalized
second-order moment of the brightest 20\% of the galaxy's flux
($\overline{M}_{20}$). In this paper, the first of a series, we
describe the technical aspects regarding the surface photometry, and
present a basic analysis of the global and structural properties of
the SINGS galaxies at different wavelengths. The homogeneity in the
acquisition, reduction, and analysis of the results presented here
makes of these data ideal for multiple unanticipated studies on the
radial distribution of the properties of stars, dust, and gas in
galaxies. Our radial profiles show a wide range of morphologies and
multiple components (bulges, exponential disks, inner and outer disk
truncations, etc.) that vary not only from galaxy to galaxy but also
with wavelength for a given object. In the optical and near-IR, the
SINGS galaxies occupy the same regions in the
$C_{42}$-$A$-$G$-$\overline{M}_{20}$ parameter space as other normal
galaxies in previous studies. However, they appear much less centrally
concentrated, more asymmetric and with larger values of $G$ when
viewed in the UV (due to star-forming clumps scattered across the
disk) and in the mid-IR (due to the emission of Polycyclic Aromatic
Hydrocarbons at 8.0\,$\micron$ and very hot dust at 24\,$\micron$). In
the accompanying paper (Mu\~noz-Mateos et al$.$ 2009) we focus on the
the radial distribution of dust properties in the SINGS galaxies,
providing a detailed analysis of the radial variation of the
attenuation, the dust column density, the dust-to-gas ratio, the
abundance of PAHs and the intensity of the heating starlight.

\end{abstract}


\section{Introduction}
Understanding the physical mechanisms that have shaped galaxies into
their present-day forms has been one of the foremost goals in
extragalactic astronomy. The current spatial distribution of stars,
dust and gas results from the joint action of different processes,
such as radially-varying gas-infall rate, star formation induced by
spiral waves, the injection of metals in the interstellar medium, and
secular re-arrangement of material within the disks, among others. In
this regard, surface photometry has proven to be a convenient way to
classify and analyze the radial structure of galaxies.

Radial profiles constitute an important observational constraint on
the predictions of theoretical models of galaxy formation and
evolution. Despite being a long-known empirical fact, the exponential
nature of spiral disks still eludes a definitive explanation. Disk
galaxies are supposed to form when baryons cool inside dark matter
halos that have formed through gravitational instability, having
initially acquired angular momentum from cosmological torques (Fall \&
Efstathiou 1980). Some authors appeal to conservation of angular
momentum during the collapse as the origin of exponential disks (see
e.g$.$ Governato et al$.$ 2007 and references therein). Others ascribe
it to secular processes, such as viscosity-driven redistribution of
angular momentum within the disks (Yoshii \& Sommer-Larsen 1989;
Ferguson \& Clarke 2001).

To further complicate this issue, van der Kruit (1979) showed that the
outer regions of spirals usually deviate from the inner exponential
profile. Indeed, we now know that purely exponential profiles seem to
be rather scarce. According to Pohlen \& Trujillo (2006), only 10\% of
nearby spirals exhibit single exponential profiles lacking any evident
change of slope\footnotemark[1] (the so-called type~I profiles). Most
disks, roughly 60\%, have an inner exponential profile followed by an
steeper outer one (type~II), and the remaining 30\% have a shallower
outer exponential (type~III). Several mechanisms have been proposed to
explain downward-bending profiles, ranging from angular momentum
cutoffs to a threshold for star formation (see Pohlen at al$.$ 2008
for a recent review on the subject).

\footnotetext[1]{This is often referred to as a `break' or
`truncation'. This can be misleading, as it suggests a sudden cutoff
in the emission, while it is just a change of slope. However, since it
has become a convention, throughout this paper we will use `truncated'
and `anti-truncated' as synonyms of `down-bending' and `up-bending',
respectively.}

Surface photometry in the ultraviolet, which traces recent ($\lesssim
1$\,Gyr) star formation, has also led to new and surprising
observational tests of our understanding of disk evolution. The
discovery of extended UV (XUV) emission in the outskirts of many
spirals (Thilker et al$.$ 2005, 2007; Gil de Paz et al$.$ 2005)
provides a first-hand view of present-day disk growth and assembly. UV
profiles can be combined with mid- and far-infrared ones to derive the
radial variation of dust attenuation in spirals (Boissier et al$.$
2004, 2005, 2007). Since both stars and dust contribute to the
observed color gradients in galaxies (de Jong 1996; MacArthur et al$.$
2004), these extinction profiles are key to disentangling the two
effects, and thereby to interpret broadband color gradients in terms
of radial changes in the mean age of the stellar populations
(Mu\~noz-Mateos et al$.$ 2007). Furthermore, when compared with gas
profiles, one can derive the radial variation of the dust-to-gas
radio, which is also expected to depend on the star formation history
at different radii (Boissier et al$.$ 2004). Besides, radial profiles
can be also used to constrain the mathematical form of the star
formation law (Wong \& Blitz 2002; Heyer et al$.$ 2004; Boissier et
al$.$ 2007; Thilker et al$.$ 2007).

Light profiles have also played an important role when quantifying the
morphology of galaxies. Ever since spirals were identified as having a
bulge and an exponential disk with different light profiles (de
Vaucouleurs 1958; Freeman 1970), bulge-disk decompositions and
S\'ersic-profile fitting (S\'ersic 1968) have been routinely used to
quantify galaxy morphology. However, the applicability of these
methods can be hampered by the presence of bars, compact nuclei,
bright spiral arms or disk truncations, or when observing at
wavelengths tracing recent star formation.

Non-parametric morphology estimators differ from the B/D ratio or the
S\'ersic index in that they do not implicitly assume a functional form
for the spatial distribution of light in galaxies. The concentration
index (de Vaucouleurs 1977; Kent 1985) and the rotational asymmetry
(Schade et al$.$ 1995) are the most extensively used parameters of
this kind in the literature (Abraham et al$.$ 1996a, 1996b; Bershady
et al$.$ 2000; Kuchinski et al$.$ 2000, 2001; Conselice et al$.$ 2000;
Taylor-Mager et al$.$ 2007). More recently, the toolbox of
non-parametric morphology estimators has been upgraded with the
incorporation of new indicators. The Gini coefficient ($G$, Abraham et
al$.$ 2003) measures the relative contribution of bright and faint
pixels to the total galaxy luminosity. The normalized second-order
moment of the pixels constituting the brightest 20\% of the galaxy flux
($\overline{M}_{20}$, Lotz et al$.$ 2004) is closely related to
classical concentration indices, but it is more weighted by the
spatial distribution of bright off-center regions.

It follows from the discussion above that a complete description of
the morphology of galaxies across a wide wavelength range is paramount
for our understanding of galaxy buildup. This is the first paper in a
series aiming to characterize the radial distribution of stars, dust
and gas in nearby galaxies, making use of the multi-wavelength data
sets available for the galaxies in the {\it Spitzer} Infrared Nearby
Galaxies Survey (SINGS; Kennicutt et al$.$ 2003). Here we present
surface brightness radial profiles ranging from the far-ultraviolet to
the far-infrared, along with asymptotic magnitudes and the
aforementioned non-parametric morphology estimators. In the
accompanying paper (Mu\~noz-Mateos et al$.$ 2009, Paper~II hereafter),
we carry out a thorough study of the radial distribution of several
dust properties: attenuation, surface density, PAH abundance, heating
starlight intensity and dust-to-gas ratio. Finally, we are also
fitting our radial profiles with models for the chemical and
spectro-photometric evolution of spirals of Boissier \& Prantzos
(2000), to infer the radial change of the SFH in a self-consistent
frame. The results of this study will be presented in a forthcoming
paper. In addition to the analysis of the basic global and structural
parameters of the SINGS sample, the data presented here possess an
important legacy value for future studies of galactic structure.

This paper is organized as follows. In Section~\ref{sample} we outline
the characteristics of the galaxy sample and the multi-wavelength
imagery. Section~\ref{analysis} details the technical aspects of the
analysis, such as removing foreground and background objects in the
original images, obtaining the radial profiles and measuring the
morphological estimators. The results are discussed in
Section~\ref{results}, and our main conclusions are
finally summarized in Section~\ref{conclusions}. In
Appendix~\ref{recalib} we explain the corrections applied to the
zero-points of some of the optical images.

\section{The sample and data}\label{sample}
The SINGS sample (Kennicutt et al$.$ 2003) consists of 75 nearby
galaxies selected to cover the range in morphological type, luminosity
and FIR/optical luminosity observed in the local universe. Moreover,
the SINGS galaxies also span a reasonably wide range in additional
properties, such as nuclear activity, spiral and bar structure,
inclination, surface brightness and environment. It lacks, however,
any significant luminous or ultra-luminous infrared galaxy (i.e$.$
with $L_{\mathrm{IR}}>10^{11}\mathrm{\,L}_{\odot}$). All galaxies are
closer than 30\,Mpc, with the median distance being 10\,Mpc. Note,
however, that early-type galaxies (E, S0 and Sa-Sab) tend to be
further away than the bulk of the sample, while irregulars are usually
much closer. In spite of being neither a flux- nor a volume-limited
sample (and thus lacking statistical power as a whole), this sample
constitutes an excellent benchmark to study the interplay between star
formation and the ISM in environments with a large variety of physical
properties. The main properties of the SINGS galaxies are summarized
in Table~\ref{sample_tab}.

Throughout the remainder of this paper we quote all the
multi-wavelength data in the AB magnitude system, otherwise
mentioned. The AB magnitudes can be translated into flux densities
following the definition of Oke (1974):
\begin{equation}
m_{\mathrm{AB}} (\mathrm{mag})=-2.5\log F_{\nu} (\mathrm{Jy})+8.9
\end{equation}

\subsection{GALEX data}
The GALEX mission (Martin et al$.$ 2005) has observed nearly all SINGS
galaxies in the FUV ($\lambda_{eff}=151.6$\,nm) and the NUV
($\lambda_{eff}=226.7$\,nm). A dichroic beam splitter allows these
observations to be carried out simultaneously at both bands, although
only NUV data are available for a few galaxies, since the FUV detector
had to be occasionally turned off for safety reasons. Photon lists are
created from individual photon detections, and are then translated
into intensity maps, with a final pixel-scale of 1.5$\arcsec$. The
flux calibration is based on white dwarf standard stars, with an
estimated uncertainty of 0.15\,mag at both wavelengths for the
pipeline version used here (the same as in Gil de Paz et al$.$ 2007).

The size of the PSF varies slightly with the position on the detector
and the brightness of the source, but the FWHM is typically
6$\arcsec$, which corresponds to a spatial scale of $\sim$300\,pc at
the median distance of the SINGS sample. The GALEX resolution nicely
matches that of the MIPS 24\,$\micron$ band, although the PSFs are
different, the Airy rings being noticeable only in the 24\,$\micron$
data.

\subsection{Optical data}\label{opt_data}
Optical images for the SINGS galaxies were taken at the Kitt Peak
National Observatory (KPNO) 2.1\,m telescope and the Cerro Tololo
Inter-American Observatory (CTIO) 1.5\,m telescope, using Harris
$BVRI$ filters. The raw frames have pixel scales of 0.305$\arcsec$ and
0.433$\arcsec$ for the KPNO and CTIO telescopes, respectively, and were
processed following standard reduction routines for optical
images. These include bias subtraction, flat-field correction,
cosmic-ray removal and mosaicking for those galaxies larger than the
instrument's field of view (10$\arcmin$ at KPNO and 14.5$\arcmin$ at CTIO).

The images were flux-calibrated with photometric standard stars imaged
during each observing run. However, the final frames exhibit
non-negligible zero-points offsets $-$compared to the global shape of
the SED$-$ whose origin is difficult to trace back and has not been
fully elucidated. In order to deal with this problem, we initially
searched for all publicly available optical images for the SINGS
galaxies using NED. After closely examining the retrieved data, we
soon realized that the data quality was quite heterogeneous in terms
of spatial coverage, image depth and spatial resolution.

We finally opted for using images from the Sloan Digital Sky Survey
(SDSS; York et al$.$ 2000) for as many galaxies as possible. We relied
on imagery from the Data Release 6 (Adelman-McCarthy et al$.$
2008). The necessary frames for each galaxy were downloaded and
mosaicked together. The conversion from counts to physical units was
carried out following the prescriptions provided in the SDSS DR6 Flux
Calibration Guide\footnotemark[2], using the calibration factors of
the frame that was used as a flux reference when building each
mosaic. Surface brightness profiles were then measured on these images
as explained in Section~\ref{surf_phot}. Although absolute zero-point
errors for the SDSS photometry have not been reported yet, the
relative photometry is known to have a uniformity of 2\%-3\%.

\footnotetext[2]{http://www.sdss.org/dr6/algorithms/fluxcal.html}

Despite their short exposure time (54 seconds), we found the SDSS data
to be surprisingly useful for measuring the profiles of the outer
regions of the galaxies. Since they are taken in drift-scan mode, they
exhibit an almost flat background, thus matching the quality of images
taken with similar telescopes but larger exposure times (see also
Erwin et al$.$ 2008 in this regard).

For those galaxies not in the SDSS, we opted for recalibrating their
SINGS optical images using the extensive catalog of aperture
photometry compiled in Prugniel \& Heraudeau (1998). The recalibration
procedure is described in detail in Appendix~\ref{recalib}. We
estimate that the zero-point error of our final recalibrated data is
10\%-15\%.

In Table~\ref{sample_tab} we indicate which approach was chosen for
each particular galaxy, namely: to measure the profiles on the SDSS
data (32 galaxies) or to recalibrate the original SINGS images with
the catalog of Prugniel \& Heraudeau (1998) (23 galaxies). The
remaining 20 objects (mostly irregular galaxies) were not included in
this catalog or, if they were, the resulting recalibrated optical
points did not look reliable enough when compared to the adjacent
photometric data-points.

It should be noted that these zero-point issues do not affect the
non-parametric morphological estimators. These structural parameters
were thus measured on both the SDSS and the SINGS images, regardless
of whether the latter had been recalibrated or not.

We do not provide neither photometry nor structural parameters in the
very few cases when significant patches of the galaxy are missing from
the mosaics (as happens, for instance, in the $R$-band image of
NGC~3031).

\subsection{2MASS data}
Since the SINGS galaxies are too large to fit into a single 2MASS
scan, the corresponding mosaics were retrieved from the 2MASS Large
Galaxy Atlas (LGA; Jarrett et al$.$ 2003). The reader is referred to
that paper for an in-depth description of the LGA data. The images
have a pixel-scale of 1$\arcsec$, and a PSF FWHM of $2\arcsec -
3\arcsec$, depending on the seeing conditions (120 pc at the median
distance of 10\,Mpc). The calibration errors are estimated to be
0.011, 0.007 and 0.007\,mag in $J$, $H$ and $K_S$, respectively (Cutri
et al$.$ 2003). The Vega-based magnitudes of 2MASS were converted into
the AB system by applying the zero-point corrections quoted in Cohen
et al$.$ (2003):
\begin{eqnarray}
J_{\mathrm{AB}}&=&J_{\mathrm{Vega}}+0.894\nonumber\\
H_{\mathrm{AB}}&=&H_{\mathrm{Vega}}+1.374\nonumber\\
K_{\mathrm{S\ AB}}&=&K_{\mathrm{S\ Vega}}+1.840
\end{eqnarray}

\subsection{Spitzer data}
Mid-infrared images at 3.6, 4.5, 5.8 and 8.0\,$\micron$ were obtained
using the Infrared Array Camera (IRAC, Fazio et al$.$ 2004) onboard
{\it Spitzer} (Werner et al$.$ 2004). The FWHM of the PSF at each channel
are 1.7$\arcsec$, 1.7$\arcsec$, 1.9$\arcsec$ and 2.0$\arcsec$
respectively, probing physical scales of 80-100\,pc at the median
distance of the sample. Mosaics were taken for galaxies larger than
IRAC's field of view ($\sim 5\arcmin$), while the smaller ones were
observed in a single dither pattern.

The images provided in the SINGS Fourth Data Delivery are based on the
Version 13 Basic Calibrated Data produced by the {\it Spitzer} Science
Center. They have undergone additional processing to account for
geometrical distortion and rotation, residual bias structure, image
offsets, bias drift, cosmic ray removal and constant background
subtraction, as well as photometric calibration. The final pixel scale
is set to 0.75$\arcsec$. Although the estimated zero-point error is
$\sim 2$\% (Reach et al$.$ 2005), the uncertainty in the aperture
corrections increase the global error up to $\sim 10$\%
\footnotemark[3]. These corrections account for the diffuse scattering
of incoming photons throughout the IRAC array, and must be applied
even for large apertures, since the photometry is normalized to finite
apertures of 12$\arcsec$, rather than infinite ones.

\footnotetext[3]{\tt http://ssc.spitzer.caltech.edu/irac/calib/extcal/}

The Multi-band Imaging Photometer (MIPS, Rieke et al$.$ 2004) was used
to image the galaxies at 24, 70 and 160$\micron$, with FWHM of
5.7$\arcsec$, 16$\arcsec$ and 38$\arcsec$, respectively. These
resolutions correspond to physical scales of 0.28, 0.78 and 1.84\,kpc
at the median distance of the sample. The observations were carried
out using the scan-mapping mode, visiting each galaxy in separate
epochs and with different orientations to identify asteroids and
remove detector artifacts at the two longer wavelengths. Further
processing by the SINGS team include conversion of 70 and
160\,$\micron$ signal ramps to slopes, flat-fielding, subtraction of
zodiacal light at 24\,$\micron$, removal of short-term variations in
the signal due to drift, background subtraction, final mosaicking and
calibration. The delivered frames at 24, 70 and 160\,$\micron$ have
pixel-scales of 1.5$\arcsec$, 4.5$\arcsec$ and 9.0$\arcsec$,
respectively, chosen to be integer multiples of that of the IRAC
mosaics, while still properly sampling the MIPS PSF. The estimated
calibration errors at each band are 4\%, 5\% and 12\%, respectively
(Engelbracht et al$.$ 2007; Gordon et al$.$ 2007; Stansberry et al$.$
2007).

\section{Analysis}\label{analysis}
\subsection{Object masking}\label{masking}
Masking the brightest field stars and background galaxies is essential
to obtain good quality surface brightness profiles. In order to detect
these sources, we run SExtractor (Bertin \& Arnouts 1996) in dual mode
on the IRAC images using the 3.6\,$\micron$ one as the detection
image. Although SExtractor is mainly oriented towards source-detection
in large-scale galaxy surveys, the configuration parameters can be
tuned so that it can also deblend and extract sources embedded within
the light of nearby galaxies. Note that we only use the results from
SExtractor to identify and mask foreground stars and background
galaxies.

For each detected source, SExtractor yields a stellarity index
(\texttt{CLASS\_STAR}) ranging from 0 for extended sources to 1 for
point-like ones. Following the methodology described in Robin et al$.$
(2007), we concluded that $\mathrm{\texttt{CLASS\_STAR}} = 0.8$
provides a sufficiently clean separation between resolved and
unresolved sources. Objects with $f_{3.6\micron}>f_{5.8\micron}$ and
$\mathrm{\texttt{CLASS\_STAR}} \geq 0.8$ are automatically classified
as foreground stars. A visual inspection confirms that these two
simple criteria are quite effective at masking most of them. In some
galaxies where individual stars might be seen (especially in the outer
regions), an additional criterion is added to avoid excessive masking
in these areas. We define a `contrast parameter' $\Gamma =
|f_{\mathrm{global}}-f_{\mathrm{local}}|/f_{\mathrm{global}}$, where
$f_{\mathrm{local}}$ is the 3.6\,$\micron$ flux density computed after
having subtracted a local background\footnotemark[4] and
$f_{\mathrm{global}}$ is obtained assuming a constant background value
all over the image, the same for all sources. In other words, $\Gamma$
is a quantitative measure of how bright is a given source compared to
its surroundings. By tuning $\Gamma$ we can control to what extent
doubtful sources embedded in the galaxy light are masked or not. We
found that sources with $\Gamma < 0.2$ are usually too contaminated by
galaxy light to allow for a proper classification. This is not a
concern, anyway, for these sources are too faint to affect our radial
profiles, asymptotic magnitudes and morphological parameters.

\footnotetext[4]{SExtractor applies a median filter in boxes with a
user-defined width (15 pixels in our case), and then performs a
bi-cubic spline interpolation to produce a background map. We opted to
use this as the local background when doing the photometry, instead of
measuring it within a certain ring around each source, since we are
interested in removing the smooth emission arising from the galaxy.}

\begin{figure}
\resizebox{1\hsize}{!}{\includegraphics{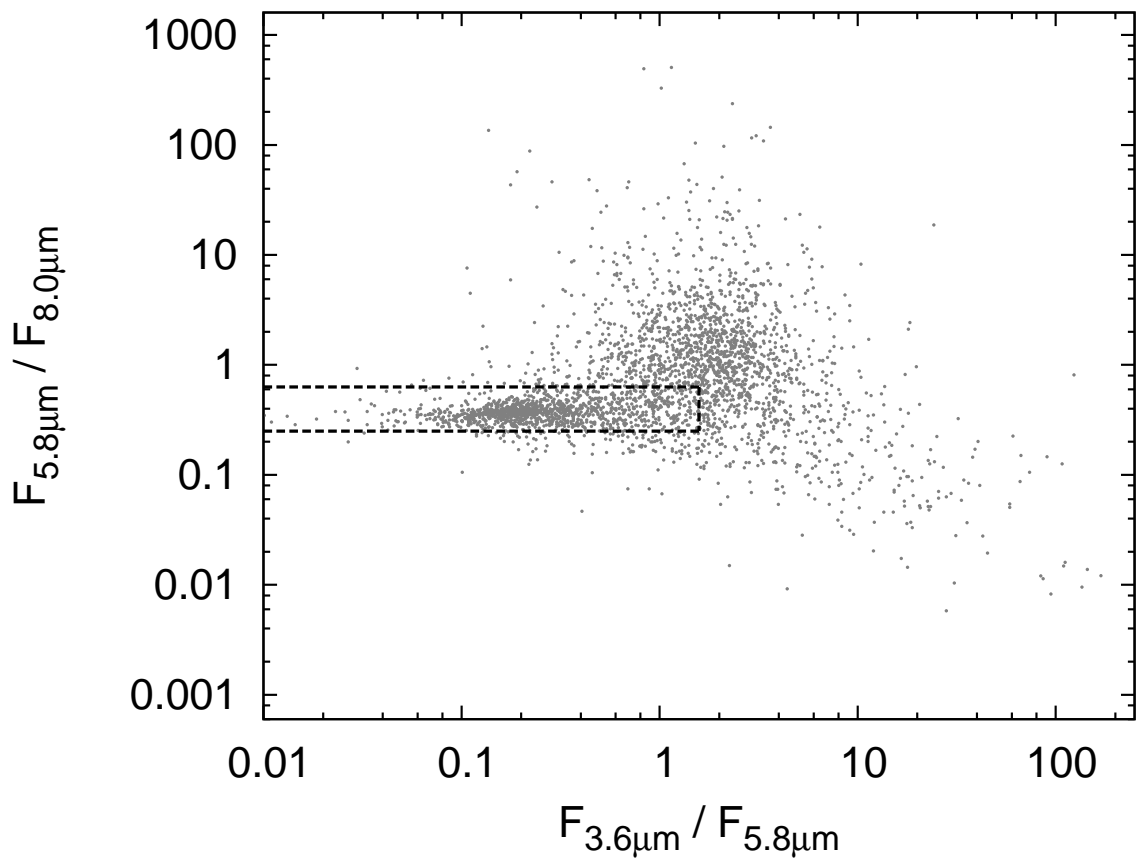}}
\caption{IRAC color-color plot of all the diffuse sources detected in
  the 3.6\,$\micron$ image of NGC~6946. Star-forming regions within
  the galaxy are clustered in a thin cloud of points with a roughly
  constant $(5.8\micron-8.0\micron)$ color. In general, local HII
  regions of the galaxies in our sample lie within the rectangular
  region shown with dashed lines. The corresponding colors (see
  Section~\ref{masking}) are used to classify extended objects either
  as local HII regions or background galaxies. The preliminary masks
  generated this way are later checked and refined by
  hand.\label{sources}}
\end{figure}

As for extended objects (i.e$.$ those with
$\mathrm{\texttt{CLASS\_STAR}} < 0.8$), they can be either local HII
regions belonging to the nearby galaxy or entire background
galaxies. The former can be easily identified thanks to the fact that
the ratio of the emission at 5.8 and 8.0\,$\micron$ due to PAHs holds
rather constant in the diffuse ISM of nearby galaxies (Draine \& Li
2007). This is clearly seen in Fig.~\ref{sources}, where we plot an
IRAC color-color diagram for all the sources detected in the IRAC
images of NGC~6946. Star-forming regions belonging to NGC~6946 are
arranged in a very thin cloud with an almost constant
$(5.8\micron-8.0\micron)$ color, while background galaxies lie outside
this region, because they are either redshifted spirals or
ellipticals. After several trials, we found that most local HII
regions in our galaxies can be isolated by means of the following
color criteria:
\begin{eqnarray}
F_{5.8\micron}/F_{8.0\micron} &>& 0.25\\
F_{5.8\micron}/F_{8.0\micron} &<& 0.63\\
F_{3.6\micron}/F_{5.8\micron} &<& 1.58
\end{eqnarray}

These criteria, combined with SExtractor's stellarity index and a
contrast parameter similar to the one described above for stars,
perform rather well for the majority of galaxies in the SINGS
sample. It should be noted that the 5.8\,$\micron$ and 8.0\,$\micron$
bands may be highly contaminated by stellar emission in the central
regions of galaxies with large bulges. Indeed, we checked that some
sources in these regions failed the color criteria, but they were not
masked thanks to the contrast parameter. Otherwise, they would have
been misclassified as background ellipticals.

\begin{figure}
\resizebox{1\hsize}{!}{\includegraphics{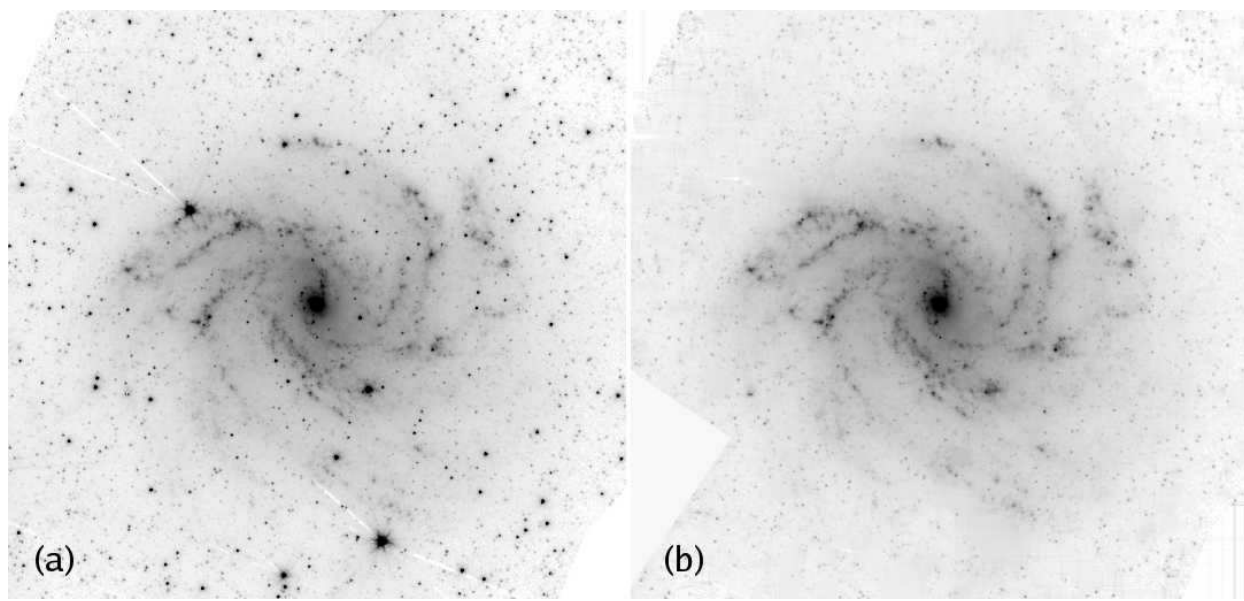}}
\caption{(a): Original 3.6\,$\micron$ image of NGC~6946. (b):
Resulting image after having detected and cleaned foreground stars,
background galaxies and artifacts. The same brightness cuts are used
to display both images. The diffraction spikes and halo emerging from
a very bright star southeast of the galaxy, although barely visible in
this printed version of the FITS image, were also masked, hence the
blank area in the bottom-left region of the
image.\label{ngc6946_clean}}
\end{figure}

This technique was just used to automatically generate masks which
were then visually inspected to detect possible errors; in the few
cases when it was needed, we unmasked regions of the galaxy that had
been misclassified as field objects, or masked sources that had eluded
the detection and classification process. These masks were then
applied to all images at every wavelength and, for each one of them,
artifacts such as bleeding, reflections and diffraction spikes were
also cleaned out. As an example, in Fig.~\ref{ngc6946_clean} we show
the original 3.6\,$\micron$ image of NGC~6946 and the one resulting
after the cleaning process.

By no means is this detection and classification procedure intended to
be accurate at the level of individual sources. Rather, it simply
allows us to mask the most relevant objects that could contaminate the
emission from the nearby galaxy when measuring the profiles.

\subsection{Surface photometry}\label{surf_phot}
We have worked with two sets of radial profiles: high resolution
profiles for all data between the FUV and 24\,$\micron$, with a radial
step of 6$\arcsec$ (matching the GALEX and MIPS 24\,$\micron$ FWHM)
and lower resolution ones with a radial increment of 12$\arcsec$ at 70
and 160$\micron$. While the latter actually oversample the MIPS PSF at
those bands, such a radial step is desirable to properly measure the
asymptotic magnitudes. In Paper~II we present 48$\arcsec$-resolution
profiles measured on GALEX, IRAC and MIPS images, spatially degraded
in order to match the size and shape of their PSFs to that of the
160$\micron$ channel. These profiles are used to determine the radial
variation of several dust properties and are also combined with HI
profiles from The HI Nearby Galaxies Survey (THINGS; Walter et al$.$
2008) and CO profiles from the literature, to study the radial
variation of the dust-to-gas ratio.

Surface brightness profiles in GALEX FUV and NUV bands were presented
by Gil de Paz et al$.$ (2007); 2MASS K-band profiles for 16 out of the
75 SINGS galaxies were included in the sample studied by
Mu\~noz-Mateos et al$.$ (2007). The profiles presented here were
obtained in the same way as was done for the GALEX and 2MASS images in
those papers. We used IRAF\footnotemark[5] task {\sc ellipse} to
measure the mean intensity along elliptical isophotes with fixed
ellipticity and position angle, equal to those of the $\mu_{B}$ = 25
mag arcsec$^{-2}$ isophote from the RC3 catalog\footnotemark[6] (de
Vaucouleurs et al$.$ 1991). For those galaxies for which these
geometrical parameters were not quoted in the RC3 catalog, we used the
values provided in NED. The center of these ellipses were set at the
coordinates shown in Table~\ref{sample_tab}, with constant increments
of 6$\arcsec$ and 12$\arcsec$ along the semimajor axis to a final
radius at least 1.5 times the D25 diameter. This value was increased
if significant emission was seen beyond that radius (especially in the
UV bands). In order to measure the different fluxes in the same
regions of each galaxy, the same set of elliptical isophotes was used
in all bands.

\footnotetext[5]{IRAF is distributed by the National Optical Astronomy
Observatories, which are operated by the Association of Universities
for Research in Astronomy, Inc., under cooperative agreement with the
National Science Foundation.}

\footnotetext[6]{Except for NGC~5194, whose original values were
highly affected by its companion galaxy, NGC~5195. Our finally adopted
values better match the actual shape of NGC~5194 itself.}

In order to derive the uncertainties in the surface brightness
profiles, we followed the methodology described in Gil de Paz \&
Madore (2005). The following expression relates the intensity in
counts per pixel ($I$) with the surface brightness in
mag\,arcsec$^{-2}$ ($\mu$):
\begin{equation}
\mu=C-2.5\log(I-I_{\mathrm{sky}})+5\log(\mathrm{arcsec\, pixel}^{-1})
\end{equation}
where $C$ is the corresponding zero-point constant. To first
order, the error in $\mu$ can be obtained as:
\begin{equation}
\Delta \mu=\sqrt{(\Delta C)^2+\left(\frac{2.5\log(e)}{I-I_{\mathrm{sky}}}\right)^2 (\Delta I^2+\Delta I^2_{\mathrm{sky}})}\label{eq_delta_mu}
\end{equation}

The uncertainty in the incident flux per pixel can be estimated assuming poissonian statistics:
\begin{equation}
\Delta I = \sqrt{\frac{I}{g_{\mathrm{eff}} N_{\mathrm{isophote}}}}\label{eq_Ilambda_error}
\end{equation}
where $g_{\mathrm{eff}}$ is the effective gain, necessary to convert
the incoming flux per pixel $I$ into electrons, and
$N_{\mathrm{isophote}}$ is the number of pixels used to compute the
mean surface brightness within each isophote. In general, a single
value of $g_{\mathrm{eff}}$ was used for all the pixels in each
frame. However, each IRAC image has an associated weight-map
indicating the coverage of each pixel by the different pointings of
the corresponding mosaic. Therefore, for the IRAC frames a spatially
varying effective gain was derived from these weight-maps.

Our approach differs from that of Gil de Paz \& Madore (2005) in the
sense that these authors rely on the $rms$ along each isophote to
estimate the error on $I$. Therefore, besides statistical
uncertainties their errorbars also include spatial variations in the
flux coming from different regions. Since our data-set spans a very
wide range of wavelengths, the $rms$ within each isophote can vary due
to the different degree of clumpiness of stars (old and young), dust
and gas. Also, in those bands where the PSF FWHM is significantly
larger than the actual size of the physical structures being probed
(e.g$.$ at the longest MIPS wavelengths), the measured $rms$ could be
artificially smoothed. Therefore, we prefer to consider only the
purely statistical error in the mean surface density within each
isophote.

As for the uncertainty in the sky level, it essentially comes from two
sources: high and low spatial frequency errors. The former results
from the combination of Poisson noise in $I_{\mathrm{sky}}$
plus pixel-to-pixel flat-fielding errors. The latter is due to
large-scale flat-fielding errors (because of residual gradients,
reflections, etc.) as well as real background structures such as
cirrus. In order to quantify these two sources of uncertainty, for
each frame we measured the sky in $\sim 20$ square regions of
$N_{\mathrm{region}}$ pixels each, randomly placed around each galaxy
far enough from it to avoid contamination from the galaxy
itself. $I_{\mathrm{sky}}$ was then determined as the mean sky
value in all boxes. We also computed the mean standard deviation,
$\langle \sigma_{\mathrm{sky}} \rangle$, and the standard deviation of
the mean sky values among different boxes, $\sigma_{\langle
\mathrm{sky} \rangle}$. Thus,
\begin{equation}
\Delta I_{\mathrm{sky}} = \sqrt {\frac{\langle \sigma_{\mathrm{sky}} \rangle^2}{N_{\mathrm{isophote}}}+\max \left(\sigma^2_{\langle\mathrm{sky} \rangle}-\frac{\langle \sigma_{\mathrm{sky}} \rangle^2}{N_{\mathrm{region}}},0\right)}
\end{equation}

The second term accounts for the large-scale background errors, and
might be especially important in the outer regions of the galaxy,
where we average the flux along widely separated regions of the
detector. Even in a frame with a perfectly flat background the
measured large-scale variance $\sigma^2_{\langle \mathrm{sky}
\rangle}$ would be nonzero. If no large-scale variations were present
and we placed several sky boxes with only one pixel each
($N_{\mathrm{region}} = 1$), then the measured large-scale variance
would be equal to the local one, $\langle \sigma_{\mathrm{sky}}
\rangle^2$. In general, $\langle \sigma_{\mathrm{sky}} \rangle^2 /
N_{\mathrm{region}}$ is the expected value of $\sigma^2_{\langle
\mathrm{sky} \rangle}$ in the absence of true background variations
across the frame. Therefore, the difference between both quantities
reflects the contribution of actual large-scale background changes to
the final error. When such a difference is negative $-$something that
can statistically happen$-$ we assume that large-scale variations are
not present.

The technique described in this section could not be fully applied to
the 70$\micron$ and 160$\micron$ images given their small size. Since
we could not fully guarantee that our sky boxes did not overlap with
the faintest isophotes of the galaxy, in these cases we relied on the
sky values and errors included in the FITS headers of each image,
which were obtained from larger mosaics.

The resulting profiles are shown in Tables~\ref{uv_opt_2mass_tab},
\ref{uv_sdss_2mass_tab} and \ref{irac_mips_tab}. The quoted errors in
the surface photometry do not include zero-point errors, which must be
considered when comparing fluxes measured in different bands. The
typical zero-point uncertainties for each band can be found in the
corresponding subsection within Section~\ref{sample}. Also, note that
the large errors in the outer regions of most profiles are usually
dominated by large-scale errors in the background determination. In
many cases there is emission from the galaxy that clearly emerges
above the local noise, so the large associated uncertainties do not
necessarily imply non-detections.

The radial profiles were corrected for Galactic extinction as in Dale
et al$.$ (2007) using the color excesses from the maps of Schlegel et
al$.$ (1998) and the extinction curve of Li \& Draine (2001), assuming
$R_{V}=3.1$.

While surface photometry with elliptical isophotes is a technique
routinely used in the literature, there are some caveats worth
mentioning. First of all, in moderately inclined galaxies with
prominent bulges or halos, such as NGC~4594 (the Sombrero Galaxy), the
shape of the RC3 ellipses is intermediate between that of the disk and
the bulge. Disentangling these components usually requires more
sophisticated procedures (see e.g$.$ Bendo et al$.$ 2006). More oblate
ellipses do certainly fit the disk in the mid- and far-IR, where it is
neatly detached from the surrounding structures. However, such
ellipses still suffer from similar problems in the optical and UV,
since they probe both the foreground half of the disk, which is
heavily obscured by dust lanes, and the background half of the disk
behind the bright bulge.

Secondly, a more general problem in all edge-on galaxies is that light
from the outer disk might end up being contaminated by more central
regions when performing the azimuthal average. Simply measuring the
surface brightness along the major axis is not devoid of problems
either, for the surface brightness at a given observed radius results
from the combined emission of sources along the line of sight, located
at very different physical distances from the center of the
galaxy. Since the optical depth depends on wavelength, a full
radiative transfer treatment is usually more suitable in these cases
(see e.g$.$ Xilouris et al$.$ 1999; Popescu et al$.$ 2000).

Finally, although irregulars may not necessarily exhibit a disk-like
structure, radial profiles are still useful as a coarse measurement of
the radial variation of their physical properties. In the case of
elliptical galaxies it should be noted that profiles with elliptical
isophotes cannot be interpreted in the same way as those of projected
disks.

\subsubsection{Corrections for the IRAC and MIPS data}

The IRAC surface brightness profiles required being corrected for
aperture effects to account for the diffuse scattering of incoming
photons throughout the IRAC array\footnotemark[7]. While usual
aperture corrections account for the extended wings of the PSF due to
the diffraction of light through the telescope optics, the ones
considered here correct for the diffraction of light through the
detector substrate (Reach et al$.$ 2005).

\footnotetext[7]{See {\tt http://ssc.spitzer.caltech.edu/irac/calib/extcal/}}

In order to properly correct our radial profiles profiles, we first
computed the growth curve, thus getting the accumulated flux up to
each given radius. We then applied the proper extended source aperture
corrections to each elliptical aperture, and then recomputed $\mu$ by
subtracting the flux of adjacent apertures.

For the 70$\micron$ image, prior to measuring the profiles we applied
the preliminary correction for nonlinearity effects quoted in Dale et
al$.$ (2007), which is derived from data presented by Gordon et al$.$
(2007).

\subsubsection{Asymptotic magnitudes}
\begin{figure}
\resizebox{!}{!}{\includegraphics{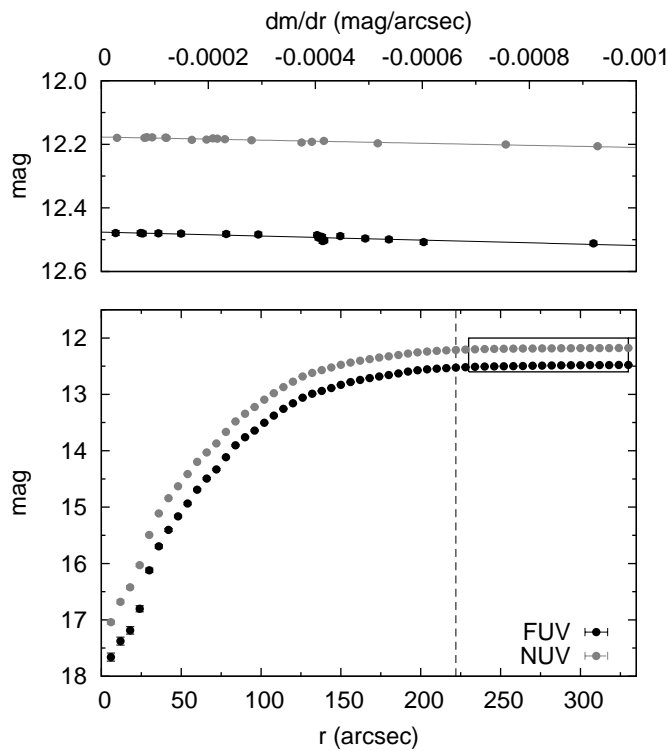}}
\caption{Sample plot showing how the asymptotic magnitudes are
derived. The bottom panel shows the growth curve of NGC~3184 in the
FUV and NUV, that is, the accumulated luminosity inside elliptical
apertures with a given radius along the semi-major axis. For
comparison, the optical size R25 is shown with a vertical dashed
line. For each data-point we compute the radial gradient of the
accumulated magnitude, $dm/dr$, and plot it against the accumulated
magnitude itself (top panel). Both quantities usually exhibit a linear
behavior in the outer regions of galaxies. The points in the upper
panel are those inside the small rectangular box in the lower
panel. A linear fit is applied to the data, and the y-intercept
(i.e$.$ the accumulated magnitude towards zero-gradient) is taken, by
definition, as the asymptotic magnitude.\label{asmag_sample}}
\end{figure}

The asymptotic magnitudes $-$that is, the ones that would be obtained
by measuring with a hypothetically infinite aperture$-$ can be derived
by means of the growth curve (see e.g$.$ Cair\'os et al$.$ 2001). The
procedure is depicted in Fig.~\ref{asmag_sample}. We computed the
radial gradient in the accumulated magnitude at each radius, which
typically exhibits a linear behavior when plotted against the
accumulated magnitude.  We then applied a weighted linear fit to the
points within a suitable outer spatial range, with the asymptotic
magnitude being the y-intercept of this fit, that is, the
extrapolation towards a zero gradient. The resulting values are shown
in Tables~\ref{asmag_uv_opt_2mass_tab}, \ref{asmag_uv_sdss_2mass_tab}
and \ref{asmag_irac_mips_tab}. The uncertainties were estimated with
the classical statistical formulae, from the residual dispersion of
the data points with respect to the fitting line. However, these
errors do not include calibration uncertainties (see
Section~\ref{sample}).

\subsection{Non-parametric morphological estimators}

\subsubsection{Concentration indices.}
From the growth curve at each wavelength, we computed the
concentration indices $C_{31}$ (de Vaucouleurs 1977) and $C_{42}$
(Kent 1985), defined as:
\begin{eqnarray}
C_{31}&=&\frac{r_{75}}{r_{25}}\\
C_{42}&=&5 \log \left(\frac{r_{80}}{r_{20}}\right)
\end{eqnarray}
where $r_x$ are the radii along the semi-major axis encompassing $x$\%
of the total flux of the galaxy at each band, the latter computed from
the asymptotic magnitude. Note that since we are dealing with
elliptical apertures, a correction for inclination needs not to be
applied. For the remainder of the analysis presented here we will focus
on $C_{42}$ only, since both indices are tightly correlated\footnotemark[8].

\footnotetext[8]{The values of $C_{31}$ can be obtained upon request from the first author.}

The resulting values of $C_{42}$ are quoted in
Table~\ref{allmorpho_tab}. We have flagged as unreliable those values
in which the inner radius $r_{20}$ is smaller than the innermost point
of our profiles (6$\arcsec$ at 24\,$\micron$ and 12$\arcsec$ at
70\,$\micron$ and 160\,$\micron$, the latter two oversampling the PSF,
see Section~\ref{surf_phot}). In these cases the quoted values are
then just lower limits. Taking into account the shape and spatial
extent of the PSF is necessary in order not to over-interpret
concentration indices in unresolved sources. This is specially
critical in the MIPS bands, given the broad and strong Airy rings of
the corresponding PSFs. We measured the concentration index $C_{42}$
on model images of the MIPS PSFs, resulting in values of 3.7, 3.5 and
3.6 at 24\,$\micron$, 70\,$\micron$ and 160\,$\micron$,
respectively. We also found that $r_{20} \simeq 2\arcsec,6\arcsec$ and
12$\arcsec$ at those bands, which lie below or close to the smallest
radius used in the galaxy profiles, so the corresponding values of
$C_{42}$ have been flagged when necessary as explained above.

Although when measuring $C_{42}$ we rely on the centers given in the
RC3 catalog, for the asymmetry and the second-order moment we follow
an iterative process to find the center (see below). We have checked
that offsets $\lesssim 3\arcsec$ have a negligible impact on
$C_{42}$. In most cases the differences between the resulting values
are just $\sim$0.01-0.001. Only in highly concentrated objects with
small values of $r_{20}$ do the discrepancies amount to $\sim$0.3-0.1.

\subsubsection{Asymmetry.}\label{asymmetry}
Several mathematical definitions for computing the asymmetry can be
found in the literature (Schade et al$.$ 1995; Abraham et al$.$ 1996b;
Kuchinski et al$.$ 2000, Conselice et al$.$ 2000), although the
philosophy behind all of them is essentially the same. It involves
comparing the original image of a galaxy with its rotated counterpart,
the rotation angle being usually $180\degr$ (although other angles can
also yield useful information, see Conselice et al$.$ 2000).

Here we adopt the definition of the asymmetry given by Abraham et
al$.$ (1996b):
\begin{equation}
A=\frac{1}{2}\left[\frac{\sum{\left| I_{180\degr}-I_0 \right|}}{\sum{\left| I_0 \right|}}-\frac{\sum{\left| B_{180\degr}-B_0 \right|}}{\sum{\left| I_0 \right|}}\right]\label{eq_asym}
\end{equation}

$I_0$ and $I_{180\degr}$ are the intensities of the original and
rotated images, respectively, and $B_0$ and $B_{180\degr}$ the
intensities of background pixels and their rotationally-symmetric
counterparts. The sum is carried out over all pixels within a certain
aperture (see below). By taking the absolute value, the sky-noise
introduces a certain positive signal in $A$ that has to be subtracted
in order to get the asymmetry of the galaxy itself. In principle this
can be done by computing the asymmetry within a reasonably large patch
of sky, free from emission coming from the galaxy (hence the second
term in Eq.~\ref{eq_asym}). However, this is not always possible in
many images of our very extended sources, and it also tends to
increase the computation time. Lauger et al$.$ (2005) demonstrate that
the noise asymmetry can be estimated by assuming poissonian statistics
for the sky:
\begin{equation}
\sum{\left| B_{180\degr}-B_0 \right|}=\frac{2}{\sqrt{\pi}}\sigma_{\mathrm{sky}}N_{\mathrm{pix}}\label{eq_sky_asym}
\end{equation}
where $\sigma_{\mathrm{sky}}$ is the sky noise, measured as explained
in Section~\ref{surf_phot}, and $N_{\mathrm{pix}}$ is the number of
pixels within the aperture used to derive the asymmetry in the
galaxy. We verified that Eq.~\ref{eq_sky_asym} is indeed valid for our
images.

The asymmetry was measured within elliptical apertures with position
angles, diameters and axis ratios equal to those of the $\mu_{B}$ = 25
mag arcsec$^{-2}$ isophotes from the RC3 or NED, and the resulting
values are shown in Table~\ref{allmorpho_tab}. The use of isophotal
radius is discouraged when comparing galaxies within large redshift
intervals, since they might be affected by cosmological surface
brightness dimming, $k$-correction, evolution and zero-point
offsets. The use of the Petrosian $\eta$-function (Petrosian 1976) is
usually more convenient in these cases (Bershady et al$.$ 2000). It is
defined as\footnotemark[9] $\eta (r)=I(r)/\langle I(r) \rangle$,
i.e$.$, the local surface brightness at a given radius $r$ divided by
the average surface brightness {\it inside} r. By definition, $\eta
(r)$ is equal to 1 at $r=0$ and approaches 0 at larger radii. The
Petrosian radius $r_P$ is such that $\eta (r_P)$ is equal to a certain
value, usually 0.2. We found that the Petrosian radius often misses
significant emission from the outer regions of many galaxies. Since
the SINGS galaxies do not suffer from the cosmological issues listed
above, the choice of employing the RC3 apertures is probably more
justified in our case. While these apertures encompass more light than
those derived from the $\eta$-function, they still miss some light of
the very outer regions of the galaxies. Increasing the aperture size
would introduce more sky in the less spatially-extended bands, thus
making the subtraction of the noise asymmetry more critical. The
choice of the RC3 ellipses constitutes a compromise solution; as long
as $A$ is computed consistently in all bands, the trends with
wavelength should remain essentially correct.

\footnotetext[9]{This is actually the inverted form of the original
Petrosian function.}

In any case, for the sake of comparison we have also computed $A$
within ellipses whose semi-major axis were set equal to the Petrosian
radius, such that $\eta (r_P)=0.2$ (see
Table~\ref{sample_tab}). Considering all bands together, we find that
the difference between the asymmetries computed with both apertures is
negligible, with $\langle A_{R25}-A_{r_P} \rangle=-0.005$ and a
scatter of $\pm 0.009$. This holds true on a band-by-band basis as
well, except in those with poorer S/N ratios (like the $u$, $z$, $J$,
$H$ and $K_S$ bands) where the offsets, while still smaller than the
scatter, seem to be statistically significant. Those slight
discrepancies are most likely due to the subtraction of the sky
asymmetry being more delicate in those cases. Conselice et al$.$
(2000) computed $A$ both within $r_P$ and $1.5 r_P$ (incidentally,
$R25 \sim 1.5 r_P$ on average for the SINGS galaxies). We have checked
from their published data that the difference between both cases is
also negligible, with a scatter similar to ours.

The center of rotation was determined by minimizing the
asymmetry. Starting from the central positions quoted in the RC3, we
computed the asymmetry over a grid of $(x,y)$ positions, recentering and
repeating the process until a minimum value was found. To account for
differences in the astrometrical calibration and resolution of the
images, we allowed a maximum difference of 3\,$\arcsec$ from the
initial central coordinates. Objects requiring larger offsets are
intrinsically asymmetric, the centers minimizing $A$ probably not
being the actual centers of the galaxy.

It should be noted that since $A$ is computed on a pixel-by-pixel
basis, the precise final values may depend on several parameters, such
as the signal-to-noise ratio and the spatial resolution. Conselice et
al$.$ (2000) showed that $A$ drops when one is unable to resolve
structures smaller than 0.5\,kpc,although for large galaxies
asymmetries with a resolution of $\sim 1$\,kpc may be still
acceptable. Bendo et al$.$ (2007) found that $A_{3.6\micron}$ does not
strongly depend on distance in the SINGS galaxies, but $A_{24\micron}$
does, thus implying that other bands also showing inherently clumpy
structures might be affected as well. We opt for a conservative
approach and flag those values of $A$ in which the spatial resolution
at each particular band does not allow resolving structures smaller
than 0.5\,kpc. This mainly affects the 70\,$\micron$ and
160\,$\micron$ bands and, in galaxies further than 17\,Mpc, the GALEX
and 24\,$\micron$ images too.

The S/N ratio can also bias the derived values of $A$ (Lotz et al$.$
2004; Lauger et al$.$ 2005). All in all, an accurate determination of
the asymmetry as a function of wavelength would require setting all
images to a common PSF size, plate scale and depth, which is beyond
the scope of this paper. Note, however, that differences in these
parameters within the SINGS imagery are not that large anyway (except
at 70\,$\micron$ and 160\,$\micron$ in terms of resolution, and also
the 2MASS bands regarding the S/N ratio). The aforementioned
corrections, therefore, are not as critical as when analyzing the
morphology of galaxies in cosmological surveys (see e.g$.$
L\'opez-Sanjuan et al$.$ 2009 and references therein). Nevertheless,
the results presented here should be treated cautiously, as our
intention is simply to depict broad general trends.

\subsubsection{Gini coefficient.}\label{gini}
The Gini coefficient (Gini 1912) is a statistical parameter widely
used in econometrics to determine how wealth is distributed in a given
population. It was adapted by Abraham et al. (2003) for galaxy
morphology classification as a proxy for the relative contribution of
bright and faint pixels to the total galaxy flux. Here we follow the
prescriptions given by Lotz et al. (2004) to compute $G$. We first
order the sky-subtracted pixels from the lowest absolute pixel
intensity to the highest one. The Gini coefficient can be then computed as:
\begin{equation}
G=\frac{1}{\overline{\left|f\right|}n(n-1)}\sum_{i=1}^n (2i-n-1)\left|f_i\right|\label{eq_gini}
\end{equation}
where $n$ is the number of pixels, $\overline{\left|f\right|}$ is the
average absolute pixel intensity and $f_i$ is the value of the pixel
$i$ once the pixels have been ranked by their absolute brightness. The
possible values of $G$ range from 0 to 1. A galaxy where the total
flux is equally distributed among all pixels would have $G=0$, and
$G=1$ would be found if one pixel was responsible of the total galaxy
flux. Generally speaking, high values of $G$ mean that most of the
galaxy flux is localized in a few pixels, whereas low values are
indicative of a more even distribution.

Although sophisticated methods like segmentation maps can be used to
define pixels belonging to the galaxy (Lotz et al$.$ 2004), here we
simply measure $G$ within the R25 elliptical apertures from the
RC3 or NED in all bands. While a precise centering is a delicate point
when computing the asymmetry, as explained in Section~\ref{asymmetry},
the Gini coefficient is not affected by this issue. However, this is a
double-edged sword, since $G$ only tells us about the relative
contribution of pixels with different intensities to the total flux,
regardless of their spatial distribution within the galaxy. In other
words: objects with entirely different morphologies may yield very
similar values of $G$ (see e.g$.$ Fig.~2 in Abraham et al$.$
2003). While the usefulness of $G$ as a standalone morphological
parameter might be somewhat limited depending on the rest-frame
wavelength of observation, its real power emerges when comparing it
with other estimators, as we will see later.

As with the asymmetry, we have analyzed the impact on $G$ of using the
Petrosian radius instead of the R25 one. As explained in Lotz et al$.$
(2004), including too many sky pixels in the aperture will tend to
increase $G$, while leaving out emission from the outer parts of the
galaxy will systematically decrease it. We have found that, on
average, $\langle G_{R25}-G_{r_P} \rangle=0.11$, with a typical
scatter of $\pm 0.10$. Since $G \sim 0.6$ for our galaxies (see
Fig.~\ref{gini_m20}), such an offset represents a relative difference
of $\sim 18$\%. By visually inspecting the images with the R25 and
Petrosian ellipses over-plotted, we have verified that the Petrosian
radius usually leaves out significant emission from the outer regions
of galaxies. The amount of missed light depends strongly on the radial
light profile, and indeed increases monotonically with light
concentration and/or the S\'ersic index (Graham et al$.$ 2005). While
$r_P$ usually encloses almost all the emission in late-type spirals,
prominent bulges and bright inner arms in earlier ones tend to
decrease $r_P$, and with it $G$ as well. Performing the measurements
within $n$ times the Petrosian radius would not help, as it would
probably introduce too many sky pixels in late-type spirals where
$r_P$ alone is already a sufficiently large aperture.

Perhaps the most flagrant case in our sample is NGC~4736, a Sab spiral
with a small bright inner ring that dominates the overall emission at
all wavelengths. The Petrosian radius at $\eta=0.2$ is 60$\arcsec$,
which is precisely the radius of the inner ring. Having a Gini
coefficient of $\sim 0.5$ within $r_P$, this galaxy would hardly
qualify as being particularly concentrated. However, the emission
actually extends much further out, with clearly visible structures
lying 350$\arcsec$ away from the center at all wavelengths. The R25
ellipse comfortably includes all this emission, and yields $G>0.8$,
which reflects more faithfully the true nature of this object.

Since the dependence of $r_P$ on light concentration may bias the
resulting values of $G$ as a function of Hubble type, and in order to
properly handle objects like NGC~4736, we have opted to keep the R25
ellipses as our measurement apertures.

The derived values of the Gini coefficient can be seen in
Table~\ref{allmorpho_tab}. We have flagged values of $G$ as unreliable
when the FWHM at a given band is larger than 0.5\,kpc at the
particular distance of each galaxy, as we did for the asymmetry. The
Gini coefficient is expected to decrease at very low signal-to-noise
ratios, so the values of $G$ measured on the 2MASS images should be
taken with care, as they could be underestimated.

\subsubsection{The second-order moment of light.}
The total second-order moment of the light in a galaxy is defined as:
\begin{equation}
M_{\mathrm{tot}}=\sum_{i=1}^n M_i=\sum_{i=1}^n f_i \left[(x_i-x_c)^2+(y_i-y_c)^2\right]\label{eq_mtot}
\end{equation}
where $f_i$ is the flux of the pixel located at $(x_i,y_i)$, and
$(x_c,y_c)$ are the coordinates of the galaxy's center. Lotz et al$.$
(2004) suggest using the normalized second-order moment of the pixels
responsible of the brightest 20\% of the total galaxy flux:
\begin{equation}
\overline{M}_{20}=\log \left(M_{20}/M_{\mathrm{tot}}\right)
\end{equation}
where $M_{20}$ is computed by ranking the pixels in order of
decreasing intensity, and then summing $M_i$ over the brightest pixels
until $\sum f_i=0.2 f_{tot}$.

The values of $\overline{M}_{20}$ are always negative. Centralized
emission yields lower (i.e$.$ more negative) values than more extended
emission. The advantage of $\overline{M}_{20}$ over the concentration
index $C_{42}$ is that since $\overline{M}_{20}$ depends on the
squared distance to the galaxy's center, it is more sensitive to the
spatial distribution of bright regions than $C_{42}$, which is usually
heavily influenced by the bulge. We measured $\overline{M}_{20}$
within the RC3 elliptical apertures, whose centers were iteratively
shifted in order to minimize $M_{\mathrm{tot}}$, as suggested by Lotz
et al$.$ (2004). These authors also found that $\overline{M}_{20}$ can
be unreliable at poor spatial resolutions, so we again impose the same
same limit of 0.5\,kpc for the physical resolution.

When comparing the values of $\overline{M}_{20}$ measured inside the
optical ellipses with those obtained within the Petrosian ellipses, we
find that $\langle \overline{M}_{20,R25} - \overline{M}_{20,r_P}
\rangle=-0.15$, with an $rms$ of $\pm 0.22$. Considering that the
second-order moment typically ranges between $-1$ and $-4$ for the
SINGS galaxies (see Fig.~\ref{gini_m20}), this offset translates into
a relative difference of just a few percent. The fact that
$\overline{M}_{20}$ is slightly smaller when measured inside the
optical ellipses is expected, since the outer regions of galaxies
between $r_P$ and $R25$ will contribute to increase $M_{\mathrm{tot}}$
but not necessarily $M_{20}$, thus decreasing the normalized moment
$\overline{M}_{20}$.

\section{Results}\label{results}

\subsection{Asymptotic magnitudes}\label{asmag}
The asymptotic magnitudes are plotted in Fig.~\ref{seds_norm} as a
function of wavelength, with galaxies grouped according to their
morphological type. The results presented here are fully consistent
with those obtained by Dale et al$.$ (2007) using aperture
photometry. Indeed our asymptotic magnitudes are in excellent
agreement with the aperture magnitudes of Dale et al$.$ (2007). We
find that the average difference
$\left|\mathrm{mag}_{\mathrm{asymp}}-\mathrm{mag}_{\mathrm{aper}}\right|$
at each band is typically below $\sim 0.07$\,mags, with a dispersion
of $\sim 0.2$\,mags.

The 4000\,\AA\ break is evident in all panels, and its amplitude
decreases towards late morphological types. The ratio of the total
infrared to UV luminosity also varies with morphological type,
reaching a maximum value in Sb-Sbc spirals (Dale et al$.$ 2007), which
can be interpreted in terms of varying attenuation (Gordon et al$.$
2000; Witt \& Gordon 2000; Buat et al$.$ 2005). This change of the
TIR-to-UV ratio is accompanied by a variation in the slope of the UV
spectra, in the sense that spirals with lower TIR-to-UV ratios also
have flatter UV spectra. This means that the (FUV$-$NUV) color can be
used as an indirect tracer of the attenuation, not only in starburst
galaxies (Calzetti et al$.$ 1994), but also in normal star-forming
galaxies, although with larger dispersion (Boissier et al$.$ 2007; Gil
de Paz et al$.$ 2007; Paper II). Note, however, that part of the UV
reddening seen in early-type disks is not entirely due to the effect
of dust, but also to their intrinsically redder old stellar
populations (Cortese et al$.$ 2008). Interestingly, most of the
variations in the SEDs for S0/a galaxies and later are mostly driven
by changes in the observed UV emission, rather than the FIR one (Dale
et al$.$ 2007). The ratio of $F_{160\micron}/F_{3.6\micron}$ varies by
a factor of 20, roughly between 10 and 200, whereas
$F_{FUV}/F_{3.6\micron}$ spans three orders of magnitude, between
0.001 and 1.

It is also worth noting that while the SEDs of elliptical and
lenticular galaxies look very similar in the optical and near-IR
range, they differ significantly in the mid- and far-IR, in the sense
that S0 galaxies tend to be more luminous in the infrared. Indeed, the
two ellipticals showing the largest infrared fluxes in
Fig.~\ref{seds_norm} seem to be rather peculiar. NGC~0855 shows
structured emission in the mid-IR, and NGC~3265 exhibits a slightly
disturbed optical morphology, and also optical emission lines
indicating some level of star formation (Dellenbusch et al$.$ 2007).

Another evident feature in these SEDs is the 8\,$\micron$ emission due
to PAHs. It is most pronounced in Sb-Sd spirals, and seems to be
almost absent in Sdm and irregulars and, to a lesser extent, in
S0/a-Sab ones. These variations can be understood in terms of
differences in the abundance of PAHs (see e.g$.$ Engelbracht et al$.$
2005, 2008 and references therein), although a detailed modeling of
the emitting properties of dust is required to properly translate
these flux ratios into chemical abundances (Draine et al$.$ 2007).

It is instructive to examine the relative contribution of the bulge
(or pseudo-bulge) and the disk to the global SEDs of these galaxies
(Fig.~\ref{seds_norm_bd}). By inspecting the FUV and 3.6\,$\micron$
profiles, we visually determined for each galaxy the radius separating
the regions dominated by the bulge and disk emission. At this radius a
sharp change in the (FUV$-$3.6\,$\micron$) color is usually seen, due
to the steeper rise of the 3.6\,$\micron$ luminosity above the main
exponential disk, as well as to the central decrease in the FUV
luminosity (see Section~\ref{profiles}). Of course, the transition
from the bulge and disk is actually gradual, so the SEDs presented
here should be just understood as bulge- or disk-dominated. Note,
however, that the mid- and far-IR within the bulge likely arises from
the disk, although circumstellar dust in the bulge itself may also
contribute.

The largest variations appear in the optical and UV bands. On average,
$F_{FUV}/F_{3.6\micron}$ in bulges appears to be one order of
magnitude fainter than in the galaxies as a whole. Of course, the
opposite behavior is seen in the disks. The mid- and far-infrared
emission, relative to the 3.6\,$\micron$ one, also seems to lie
systematically below the median global SEDs in the central regions
than in the outer ones. In other words, both the UV emission
associated with recent star formation and the infrared light arising
from dust are more radially extended than the 3.6\,$\micron$
luminosity, which probes the underlying old stellar population. This
will be more clearly seen in the Section~\ref{morpho_lambda}, where we
will analyze the variation of the concentration index from the FUV to
the FIR.

\subsection{Radial profiles}\label{profiles}
The radial profiles of the SINGS galaxies show a wide range of
morphologies, with multiple components such as bulges and
pseudo-bulges, exponential disks, inner and outer disk truncations,
antitruncations, etc. These features do not only vary among galaxies
but also with wavelength for the same object. All the multi-wavelength
radial profiles for the SINGS galaxies are shown in
Fig.~\ref{ngc3031_allprofs_offset}. Only the ones for NGC~3031 (M~81)
are included in the printed version of the journal (see the on-line
edition for the whole figure set), but they provide an overall glimpse
of the different structures that are usually discerned in the profiles
of most spiral galaxies.

The profiles of NGC~3031 in the far- and near-UV show a sharp inner
cutoff at $\sim 200\arcsec$, which imposes an upper limit of $\sim
1$\,Gyr for the last epoch of substantial star formation in the
central part of the disk, since that is the typical lifetime of stars
dominating the UV emission. These inner-truncated disks appear to be
rather common in early-type spirals and could result from gas
exhaustion in the central regions of these galaxies. These features
warn against using a simple exponential profile when performing
bulge-disk decompositions, as the disk might not extend to the very
center of these galaxies (at least not in a disk populated by stars
younger than a few Gyr). Indeed, based on optical surface brightness
profiles, Kormendy (1977) already proposed the use of such
inner-truncated exponential functions, and Baggett et al$.$ (1998)
applied that idea to a larger sample of galaxies. The (FUV$-$NUV)
color exhibits an interesting behavior in the bulge. It initially
becomes redder as we get closer to the center, from
$(\mathrm{FUV}-\mathrm{NUV})=0.6$ at $r=200\arcsec$ to 1.8 at
$r=50\arcsec$. However, for $r<50\arcsec$ the UV color gets
progressively bluer again, reaching $(\mathrm{FUV}-\mathrm{NUV})=1$ at
the center. This could be due to the so-called UV-upturn, that is, the
flux increase from $\lambda \simeq 2000$\AA\ to $\lambda \simeq
1200$\AA\ seen in the UV spectra of old stellar populations (see
e.g$.$ O'Connell 1999 and references therein). Indeed, E/S0 galaxies
usually get increasingly bluer in the UV towards the center (Ohl et
al$.$ 1998). This bluer UV emission is thought to arise from low-mass,
helium-burning stars at the end of the horizontal branch, since these
stars usually have thin envelopes that leave their hot cores
exposed. Note that M~81 harbors an active galactic nucleus (Peimbert
\& Torres-Peimbert 1981), whose influence in the UV color cannot be
discarded, although its effects $-$if present$-$ would be probably
limited to the very center of the galaxy (Marcum et al$.$ 2001).

In the optical and near-infrared, the bulge profile smoothly merges
with the inner exponential disk, which is followed by a secondary,
downward-bending exponential beyond $\sim 500\arcsec$ in the case of
NGC~3031. This is the most common type of truncation in spiral disks
(Pohlen \& Trujillo 2006). The truncation is sharpest in the UV, and
becomes progressively smeared out in the optical and near-IR. Inside
the truncation radius, the disk scale-length is larger in the UV than
in the near-IR, thus leading to a blueing with increasing radius, as
is expected from an inside-out growth scenario of disk formation (see
e.g$.$ de Jong 1996; Boissier \& Prantzos 2000; Mu\~{n}oz-Mateos et
al$.$ 2007). However, beyond the truncation radius the trend reverses,
with the stellar populations getting redder as the galactocentric
distance increases. This change in the color gradient has been indeed
found to be a common feature in truncated disks (Bakos et
al. 2008). The behavior seen in Fig.~\ref{ngc3031_allprofs_offset} is
in agreement with the predictions of the models of R\u{o}skar et
al. (2008), who argue that the outer exponential could be populated by
old stars formed closer to the center of the galaxy, and later
scattered outwards by spiral arms.

While the change of slope in the profiles of NGC~3031 is subtle at the
3.6\,$\micron$ and 4.5\,$\micron$ bands, it becomes more pronounced at
5.8\,$\micron$ and 8.0\,$\micron$, which probe the emission arising
from PAHs. The break gets even sharper at the 24\,$\micron$ band,
dominated by hot dust emission, and then becomes less pronounced at
70\,$\micron$ and 160\,$\micron$, although the blurring effect due to
the increasingly large PSF must be also taken into account. When
comparing the FIR profiles measured on the IRAC and MIPS images after
matching their PSFs with that of the 160\,$\micron$ band, we verified
that the break was still sharpest at 24\,$\micron$, meaning that the
observed variations in the break sharpness cannot be entirely due to
resolution effects. Note that in terms of the observed flux, the
infrared profiles shown here do not only depend on the radial
distribution of dust itself, but also on that of the heating sources
(see Paper~II).

Finally, there is one last feature in the profiles of NGC~3031 that
only shows up in the FUV and NUV bands: an extended, shallower
component beyond $\sim 800$\arcsec. The GALEX images reveal structured
UV emission in the outermost regions, and this galaxy has been indeed
classified as an extended UV disk (Thilker et al$.$ 2007), a
phenomenon first discovered in M~83 (Thilker et al$.$ 2005) and
NGC~4625 (Gil de Paz et al$.$ 2005). These extended components are
most easily detected in the UV partly because of the higher surface
brightness sensitivity at those wavelengths. Underlying stellar
emission may be seen at other wavelengths, although the UV-nIR colors
tend to be blue (Thilker et al$.$ 2007). The XUV emission in NGC~3031
seems to be tidally structured, as a result of the interaction with
other galaxies belonging to the M~81 group. However, in many other
XUV-disks the organized filamentary UV emission does seem to arise
from the outward propagation of spiral waves, instead of tidal
interactions. In this sense, Bush et al$.$ (2008) argue that such a
phenomenon can take place in spirals with pre-existing extended HI
disks without the need of invoking the presence of a companion.

\subsection{Morphology}
Quantifying galaxy morphology at different wavelengths is not only
useful to better understand the spatial distribution of the different
components of galaxies, but also to automate morphological
classification in modern surveys (see e.g$.$ Scarlata et al$.$
2007). Most of the work that has been done in this field has been
carried out in the optical range, although some authors have also
considered the UV range (Kuchinski et al$.$ 2000, 2001; Burgarella et
al$.$ 2001; Lauger et al$.$ 2005; Taylor-Mager et al$.$ 2007). These
authors found that galaxies generally exhibit larger asymmetries and
lower concentration indices in the UV than in the optical, since the
UV traces recent star formation, which has a clumpier and more
radially extended spatial distribution than the intermediate-age stars
seen in the optical. These considerations are particularly important
for comparisons with samples of galaxies at $z \sim 1-2$, as optical
observations of these objects probe the restframe UV.

With the advent of {\it Spitzer}, similar analysis are now possible in
the IR as well. Bendo et al$.$ (2007) computed non-parametric
estimators on the 3.6\,$\micron$ and 24\,$\micron$ images of the SINGS
galaxies, finding that the 24\,$\micron$ emission arising from very
hot dust is usually more extended and asymmetric in late-type galaxies
than in early-type ones, even when the 3.6\,$\micron$ data are
degraded to the same resolution as the 24\,$\micron$ data.

In this section we provide non-parametric morphological estimators for
the SINGS galaxies from the FUV to the FIR. While an exhaustive
analysis of the morphological implications of these results is beyond
the scope of this paper, a brief discussion is also presented.

\subsubsection{Morphology as a function of wavelength}\label{morpho_lambda}

\paragraph{Concentration index.}
In Fig.~\ref{conc_asym} we show how $C_{42}$ varies with wavelength
for galaxies of different morphological types. Dashed lines are used
when $r_{20}$ is smaller than the innermost point of our profiles
(6$\arcsec$ at 24\,$\micron$ and 12$\arcsec$ at 70\,$\micron$ and
160\,$\micron$). In order to detect possible biases due to the spatial
resolution, we show red (blue) lines with the median values for the
galaxies further (closer) than the median distance of the galaxies
within each morphological bin. These median distances were
individually computed for each bin of Hubble types, since late-type
galaxies in the sample are closer on average than early-type ones. The
values are 17\,Mpc (E-S0), 17\,Mpc (S0/a-Sab), 15\,Mpc (Sb-Sbc),
9\,Mpc (Sc-Sd) and 4\,Mpc (Sdm-Irr).

Ellipticals and lenticulars exhibit $C_{42} \gtrsim 3$ at all
wavelengths, but spirals present a much more varied behavior. Their
concentration indices typically lie below 3 in the UV, due to the
contribution of young stars spread across the disk. Then it suddenly
rises in the optical, as the bulge dominates the overall light
concentration longward of the Balmer break ($\lambda \sim
0.36\micron$). Early-type spirals usually show central depletions in
their UV emission, with inner-truncated disks, while in the optical
and near-IR the bulge is more prominent, hence the large discontinuity
between $C_{42}(\mathrm{UV})$ and $C_{42}(\mathrm{opt})$. Such a jump
can be enhanced by the presence of rings (Lauger et al$.$ 2005). The
difference becomes progressively less pronounced in later Hubble
types, as the transition from the central bulge (or pseudo-bulge) and
the disk becomes more gradual.

The concentration index rises slowly from the optical to the near-IR,
and then drops again at 5.8\,$\micron$ and 8.0\,$\micron$, since these
bands probe the spatial distribution of PAHs. The amplitude of this
`break' in $C_{42}$ decreases in late-type spirals, but is still
sharper than the one between the UV and the optical bands. While Sc
spirals and later usually exhibit low values of $C_{42}$, typically
below $\sim 3$, there is much more dispersion in galaxies or earlier
types towards larger concentration indices. By analyzing the
24\,$\micron$ morphologies of the SINGS galaxies, Bendo et al$.$
(2007) suggested that bars might increase light concentration at
24\,$\micron$ by enhancing nuclear star formation activity, although
the statistical evidence was not compelling.

There are some galaxies in each panel having significantly larger
infrared concentration indices than the remaining galaxies of the same
morphological type. This is the case of NGC~1291 (S0/a), NGC~1512
(Sa), NGC~1097 (Sb), NGC~3351 (Sb) and NGC~4536 (Sbc). These galaxies
exhibit signs of very intense nuclear and/or circumnuclear emission
when inspected at 8\,$\micron$ and 24\,$\micron$, hence their
unusually large concentrations. Note that although NGC~1291 has an
outer ring, the nuclear emission dominates the total luminosity at
24\,$\micron$ and 70\,$\micron$, hence the large values of $C_{42}$ at
these bands.

The galaxy exhibiting high values of $C_{42}$ in the panel for Sc-Sd
spirals is NGC~5033. Unlike the galaxies mentioned above, which depart
from the general trends only in the infrared, NGC~5033 is above all
Sc-Sd galaxies already in the optical. A visual inspection shows no
sign of any intense nuclear emission, but reveals however a rather
prominent bulge for an Sc spiral. The multiband radial profiles show
indeed that the light distribution of this galaxy has a non-negligible
contribution from the central bulge, which is absent in other galaxies
of the same morphological bin. The fact that this is a Seyfert galaxy
might contribute to its high concentration in the infrared.

The results presented here agree with those presented in other papers
in the literature. As an example, Taylor-Mager et al$.$ (2007)
performed a quantitative analysis of the rest-frame UV and optical
morphology of 199 nearby galaxies, using imagery from GALEX, HST and
ground-based telescopes. They found that ellipticals and lenticulars
typically have $C_{42}\simeq 4$, they same average value obtained
here. Although these authors noted a drop-off in $C_{42}$ shortward of
the Balmer break in E-S0 galaxies $-$a feature that is absent here$-$,
they warned against over-interpreting this feature, given the low S/N
of those red galaxies in the UV. They do find a clear trend with
wavelength in Sa-Sc spirals, with $C_{42}$ rising from $\sim 2$ in the
FUV to $\sim 3$ in the $I$ band. The trend we observe is fully
consistent with theirs, although our finer bins in morphological types
reveal that the wavelength dependence of $C_{42}$ actually varies
within Sa-Sc galaxies, being less pronounced in later types. Our
typical concentration indices for irregulars agree well with those of
Taylor-Mager et al$.$ (2007) (roughly 2.5). While these authors study
in great detail the morphology of mergers, such an analysis is not
possible here, since there are not major mergers in our sample and the
few galaxies with low mass companions have those objects deliberately
masked out.

\paragraph{Asymmetry.}
In Fig.~\ref{conc_asym} we also show the dependence of the asymmetry
on wavelength. Dashed lines indicate that the FWHM at a given
wavelength corresponds to a physical size larger than 0.5\,kpc at the
distance of each galaxy. As expected, ellipticals and lenticulars
exhibit low asymmetries at all wavelengths, typically below 0.15. In
S0/a galaxies and later there is a clear trend with wavelength: $A$
reaches a maximum in the FUV, where the emission is dominated by
recent star formation. As we move towards longer wavelengths in the
optical and near-IR, the intermediate-age and old stars from the bulge
and halo $-$which are more smoothly distributed across the disk than
young ones $-$ progressively decrease the global asymmetry of the
galaxies. PAHs make their appearance at 5.8\,$\micron$ and
8.0\,$\micron$, increasing the asymmetry again. Hot dust associated
with clumps of star formation tend to increase $A$ even more at
24\,$\micron$. This U-like shape of the $A_{\lambda}$ distribution is
enhanced as we progress along the Hubble sequence: intermediate- and
late-type spirals become progressively more asymmetric in the
UV-optical range, as well as in the mid-IR. Note that the increase in
asymmetry in the UV and mid-IR bands is the opposite of what would be
expected from the degradation of the PSF, thus reassuring that these
changes in $A$ are real. However, the drop-off seen at 70\,$\micron$
and 160\,$\micron$ for some galaxies is most likely the result of the
poorer resolution.

The outlier in the panel for Sc-Sd galaxies is NGC~5474. This
companion of M~101 (Drozdovsky \& Karachentsev 2000) shows a strongly
disturbed morphology, with a large plateau shifted southward with
respect to the bright central disk, hence its large
asymmetry. Interestingly, NGC~5474 looks like a normal spiral when
viewed in the UV, thus not showing much departure from the typical
asymmetries in that spectral range.

Our multi-wavelength asymmetries seem to be consistent with other
published values for nearby galaxies. We take again as an example the
work of Taylor-Mager et al$.$ (2007). As expected, they also conclude
that E-S0 galaxies exhibit the lowest asymmetries. In Sa-Sc spirals,
their asymmetries decrease from $\sim 0.8$ in the FUV to $\sim 0.2$ in
the $I$-band. Once divided by 2 $-$these authors do not include the
$1/2$ factor when computing the asymmetry$-$, these values nicely
match the ones we have derived. The agreement is not as good in very
late-type spirals and irregulars, our UV asymmetries being a bit
larger in the UV, although with considerable dispersion.

Most galaxies show a small bump in the asymmetry at 3.6\,$\micron$ and
4.5\,$\micron$. This feature becomes more evident in the latest Hubble
types, and is most likely due to the large number of background
sources detected in these bands. While our masking procedure
eliminates sources that could potentially contaminate our profiles and
magnitudes (see Section~\ref{masking} and Fig.~\ref{ngc6946_clean}),
these ubiquitous faint sources may still have a substantial impact on
the asymmetry. This effect is expected to become more noticeable in
galaxies with lower average surface brightness within the aperture
used to compute the asymmetry (usually late-type ones). We do observe
indeed a marked correlation between $A_{3.6,\ 4.5\micron}$ and the
average surface brightness at those bands, a correlation that is not
seen at other wavelengths (not shown).

\paragraph{Gini coefficient}
In Fig.~\ref{gini_m20} we show how the Gini coefficient (and also
$\overline{M}_{20}$) varies with wavelength and morphological type. As
with the asymmetry, dashed lines are used when the FWHM does not allow
resolving structures smaller than 0.5\,kpc at the distance of each
galaxy. In general, $G$ tends to decrease from early- to late-type
galaxies, especially in the optical and near-IR bands. However, the
scatter is large, and the overlap between the values of $G$ between
adjacent bins of $T$ type is considerable. No clear trend with
wavelength can be inferred for galaxies earlier than Sab, but Sb
spirals and later $G$ apparently exhibits a mild wavelength
dependence, in the sense that it seems to be larger in the UV and
mid-IR (especially at 24\,$\micron$) than in the optical and
near-IR\footnotemark[10]. What is the reason behind this modest trend
with wavelength, despite the evident change in morphology among the
bands considered here? In the optical and near-IR, pixels in the bulge
contribute to a relatively large fraction of the total galaxy
luminosity, the Gini coefficient being largest in early-type galaxies
than in late-type ones. The morphology in the UV and the mid-IR is
considerably different, yet the total galaxy luminosity can be still
dominated by emission arising from a few pixels, but this time
localized mainly in star-forming regions. As a result, $G$ can reach
values similar to or even larger than those in the optical and
near-IR. Therefore, $G$ is only expected to be correlated with central
light concentration in the optical and near-IR (see
Section~\ref{comparison_estimators}).

Our Gini coefficients seem to be consistent with those in the
literature. At 3.6\,$\micron$ and 24\,$\micron$, the agreement between
our values and those derived in Bendo et al$.$ (2007) is
excellent\footnotemark[11]. The average offset $\langle
G_{Bendo}-G_{ours} \rangle$ is jsut $-0.03$ at 3.6\,$\micron$ and
0.008 at 24\,$\micron$, the scatter being $\sim 0.03$ in both
cases. In the optical, for the galaxies we have in common with Lotz et
al$.$ (2004) our values of $G$ are, on average, $\sim 15$\%
larger. This is most likely attributable to differences in the sizes
of the regions used to perform the measurements. Lotz et al$.$ compute
$G$ within segmentation maps whose extent depends on the Petrosian
radius at $\eta=0.2$, while we do it within the R25 ellipses,
including emission that is missed by the Petrosian aperture. As
discussed in Section~\ref{gini}, this can easily lead to systematic
differences of 15-20\% in the resulting values of $G$.

\footnotetext[10]{Note that the values of $G$ in the 2MASS bands could
be underestimated, due to the poorer S/N ratio of those images
compared to the rest of the data.}

\footnotetext[11]{Note that Bendo et al$.$ matched the PSF and the
plate-scale of the 3.6\,$\micron$ images to those of the 24\,$\micron$
ones.}

\paragraph{Second-order moment.}
Not surprisingly, the behavior of $\overline{M}_{20}$ is remarkably
similar to that of $C_{42}$ (see Fig.~\ref{conc_asym}). Ellipticals
and lenticulars exhibit $\overline{M}_{20}\sim-2.5$ across the whole
wavelength range, which is indicative of highly-concentrated light
distributions. On the contrary, most Sdm galaxies and irregulars tend
to have $\overline{M}_{20} \gtrsim -1$. In spirals,
$\overline{M}_{20}$ remains high in the UV, where most of the emission
emerges from star-forming regions distributed across the
disk. Longward of the Balmer break, $\overline{M}_{20}$ decreases
abruptly, the decrement being more pronounced in early-type spirals
than in late-type ones due to the larger bulge-to-disk ratios of the
former. A second break happens in the mid-IR, where
$\overline{M}_{20}$ increases again due to the emission of PAHs at 5.8
and 8.0\,$\micron$, and hot dust at 24\,$\micron$.

Our values of $\overline{M}_{20}$ agree well with published ones, both
in the optical for the galaxies in common with Lotz et al$.$ (2004),
and at 3.6\,$\micron$ and 24\,$\micron$ in the study by Bendo et al$.$
(2007).

\subsubsection{Trends between the different morphological estimators}\label{comparison_estimators}

It is apparent from Figs.~\ref{conc_asym} and \ref{gini_m20} that the
most dramatic changes in $C_{42}$ and $\overline{M}_{20}$ with Hubble
type take place at the optical and near-IR. An obvious correlation
shows up when plotting these morphological estimators at
3.6\,$\micron$ against the morphological type (Figs.~\ref{cindex_mag}a
and c). In Fig.~\ref{cindex_mag}b we plot the concentration index as a
function of the absolute magnitude in that band, which is a proxy for
the stellar mass of the galaxies. While both parameters are correlated
for galaxies fainter than -21\,mags, the trend breaks down at larger
luminosities, since galaxies with $M_{3.6\micron} \sim -22$ are
ellipticals, lenticulars and early-type spirals, the former being
typically more concentrated than spirals with similar stellar
masses. A very similar result was obtained by Boselli et al$.$ (1997)
for Virgo galaxies. These authors found a somewhat tighter trend
between $C_{42}$ and the $K$-band magnitude at the low luminosity regime,
since their sample better probed that luminosity range. While the
SINGS galaxies are quite homogeneously distributed in terms of
morphological types, the distribution in near-IR luminosity is more
peaked around relatively bright objects. The trend between
$\overline{M}_{20}$ and the absolute magnitude does not exhibit the
upward bending seen in panel (b) with the concentration index, since
both parameters are not linearly correlated, as we shall see below.

In Fig.~\ref{conc_vs_m20} we can see that at 3.6\,$\micron$, and also
in the optical bands, $C_{42}$ and $\overline{M}_{20}$ are very
tightly correlated, the slope of the trend increasing from late- to
early-type spirals. We have used the boundary-fitting code of Cardiel
(2009) to fit third-order polynomials to the upper and lower envelopes
of the data-point distribution at 3.6\,$\micron$. These curves are
replicated in all panels for the ease of comparison. The trend
presents an elbow located at $\overline{M}_{20} \sim -2, C_{42} \sim
3$, in good agreement with other works in the literature (Lotz et
al$.$ 2004; Scarlata et al$.$ 2007). The different slopes at both
sides of the elbow can be attributed to the varying contribution of
the bulge and disk to the overall light distribution (Scarlata et
al$.$ 2007). Indeed, most points leftwards of the elbow correspond to
Sc spirals and later, whereas those at larger concentration values
correspond to Sbc spirals and earlier. Since this trend is shaped by
the varying B/D ratio, it also holds in the optical bands, which still
trace the underlying stellar population, but it breaks down in the UV
and the mid-IR, which trace recent star formation. Both indicators
show that at these wavelengths galaxies exhibit much more extended
light distributions.

In Fig.~\ref{gini_vs_m20} we compare the Gini coefficient and
$\overline{M}_{20}$. We have fitted two straight lines to the
boundaries of the data-point distribution at 3.6\,$\micron$. A clear
correlation is seen in the optical and near-IR, in agreement with
previous studies (Abraham et al$.$ 2003; Lotz et al$.$ 2004). The
interpretation of $G$ as a proxy for light concentration is valid in
these bands simply because bright pixels are mainly located in the
bulge. Therefore, to first order the relative contribution of bright
pixels to the total luminosity closely depends on the bulge-to-disk
ratio, which in turn is correlated with central light
concentration. But this is no longer valid as soon as we peer into the
UV and mid-IR regimes, where galaxies usually have larger values of
$G$ and $\overline{M}_{20}$. This time, most bright pixels are
associated with star-forming regions, so $G$ is not correlated with
central concentration any more.

Interestingly, when viewed at 24\,$\micron$ normal galaxies occupy the
same region in the $G$-$\overline{M}_{20}$ plane as the ultra-luminous
infrared galaxies (ULIRGs) of Lotz et al$.$ (2004) in the $R$
band. This was already noted by Bendo et al$.$ (2007), and we confirm
that this also occurs in the FUV and 8\,$\micron$, although to a
lesser extent. The same behavior is seen in the $G$-$A$ plane (not
shown), where normal galaxies exhibit high asymmetries and Gini
coefficients at 24\,$\micron$, similar to those of ULIRGs in the
optical. However, this should not be blindly generalized. For
instance, the $R$-band data for ULIRGS in Lotz et al$.$ (2004) fill
the upper-left region of the $C_{42}$-$\overline{M}_{20}$ plane, but
our normal galaxies occupy a different area of the plot when observed
at 24\,$\micron$ (Fig.~\ref{conc_vs_m20}).

The relation between the concentration index and the asymmetry is
presented in Fig.~\ref{conc_vs_asym}. The dynamic range in $C_{42}$
and $A$ is largest in the FUV, where different morphological types
clearly delineate a sequence in the $C_{42}$-$A$ plane, although with a
certain overlap. In the optical and near-IR bands not only the dynamic
range is smaller, but also the relative arrangement of data-points
differs. For instance, Sb-Sbc spirals appear to be of later Hubble
types when viewed in the FUV, but merge with S0/a-Sab galaxies at
longer wavelengths. Note that, as stated above, the systematic upward
shift in the asymmetries of Sdm-Irr galaxies at 3.6\,$\micron$ is
likely due to background sources. In the mid-IR the $C_{42}$-$A$ sequence
lengthens again, the main difference with respect to the FUV being the
presence of sources with high central concentrations, due to intense
(circum)nuclear emission.

Unlike in the previous diagrams, where galaxies shift to an entirely
different locus in the parameter space when observed in the UV or
mid-IR, in the $C_{42}$-$A$ plane they apparently move along roughly the
same diagonal sequence. At 24\,$\micron$ there are a few points with
slightly larger asymmetries at fixed $C_{42}$ in comparison with the
optical bands, but in general the displacement takes place
diagonally. In Fig.~\ref{conc_vs_asym_all} we have plotted all
data-points for different bands simultaneously. Besides the six bands
shown in Fig.~\ref{conc_vs_asym}, we have added the NUV as well for a
better wavelength sampling. The bulk of the data-points lie within the
boundaries delineated by the dashed lines, regardless of the
wavelength. Many ULIRGs would probably lie in the upper-right region
of the plot.

\section{Conclusions}\label{conclusions}
We have obtained surface brightness profiles for the SINGS galaxies
all the way from the FUV to the FIR, in order to map the radial
structure of stars and the interstellar medium. The profiles were
measured on UV images from GALEX, optical data from KPNO, CTIO and
SDSS, near-IR images from 2MASS, and mid- and far-IR data from the
IRAC and MIPS instruments onboard {\it Spitzer}. The multi-wavelength
data-set released here may be used in a variety of galactic structure
studies beyond those presented in this paper, including morphological
classification, bulge-disk decomposition, analysis of disk truncations
or comparison with the output of disk-formation models, among others.

From the growth curves we have computed asymptotic magnitudes at
different wavelengths. The resulting SEDs lead to similar results as
those found by Dale et al$.$ (2007) using aperture photometry. The
4000\,\AA\ break is seen to decrease from early- to-late type
galaxies, reflecting changes in the mean age of the underlying stellar
populations. The total infrared-to-UV ratio, which can be used as a
proxy for the UV attenuation, also varies with morphological type,
being maximum in Sb-Sbc spirals. The 8\,$\micron$ feature associated
with PAHs stands out in Sb-Sd galaxies, and is less prominent in
earlier and later types, especially in Sdm and Im galaxies.

We have also analyzed the wavelength dependence of four non-parametric
morphological estimators: the concentration index, the asymmetry, the
Gini coefficient and the normalized second-order of the brightest 20\%
of the galaxy's flux. Ellipticals and S0s exhibit very small
asymmetries, large concentration indices and small (i.e. very
negative) values of $\overline{M}_{20}$ across the whole spectral
range considered here. Disk-like galaxies, however, display larger
variations, indicating the presence of several stellar and dust
components with different spatial distributions. Localized
star-forming complexes dominate the FUV emission, giving galaxies a
clumpy appearance at this band, where the asymmetry is therefore
maximum. Since star-formation is usually widespread across the whole
disk, the concentration index exhibits a minimum value in the
FUV. Older stellar populations are arranged in a more uniform disk and
a central bulge, thus decreasing the asymmetry and increasing the
concentration index in the optical and near-IR bands. This trend is
reversed when PAHs show up at 5.8\,$\micron$ and 8.0\,$\micron$, since
galaxies again exhibit a patchier and more radially extended
appearance. The same applies to the 24\,$\micron$ band, dominated by
hot-dust emission.

In the optical and near-IR, the Gini coefficient is correlated with
light concentration. Galaxies with more centrally concentrated light
distributions have most of their flux emerging from a few pixels, and
exhibit therefore high values of $G$. However, this trend does not
hold in the UV and mid-IR, where galaxies exhibit low concentration
indices yet high Gini coefficients, since the few pixels dominating
the total emission are now distributed all over the disk.

The data-set presented in this paper serves as the foundation for
other ongoing and more detailed studies we are carrying out. In the
accompanying paper (Mu\~noz-Mateos et al$.$ 2009) we provide an
in-depth analysis of the radial distribution of dust properties in the
SINGS galaxies, ranging from attenuation to dust column density, as
well as PAH abundance, dust-to-gas ratio and the properties of the
heating sources. These profiles are also being used to constrain the
predictions of the disk-evolution models of Boissier \& Prantzos
(2000). The results will be presented in a forthcoming paper, focusing
on the radial variation of the star formation history and its
connection with the inside-out assembly of disks.

\acknowledgments JCMM acknowledges the receipt of a Formaci\'on del
Profesorado Universitario fellowship from the Spanish Ministerio de
Educaci\'on y Ciencia. JCMM, AGdP, JZ, PGP and JG are partially
financed by the Spanish Programa Nacional de Astronom\'{\i}a y
Astrof\'{\i}sica under grant AYA2006-02358. AGdP is also financed by
the MAGPOP EU Marie Curie Research Training Network. We also thank the
anonymous referee for very useful comments that have significantly
improved the paper.

GALEX (Galaxy Evolution Explorer) is a NASA Small Explorer, launched
in 2003 April. We gratefully acknowledge NASA's support for
construction, operation, and science analysis for the GALEX mission,
developed in cooperation with the Centre National d'\'Etudes Spatiales
of France and the Korean Ministry of Science and Technology. This work
is part of SINGS, the {\it Spitzer} Infrared Nearby Galaxies
Survey. The {\it Spitzer} Space Telescope is operated by the Jet
Propulsion Laboratory, Caltech, under NASA contract 1403.

Funding for the SDSS and SDSS-II has been provided by the Alfred
P. Sloan Foundation, the Participating Institutions, the National
Science Foundation, the U.S. Department of Energy, the National
Aeronautics and Space Administration, the Japanese Monbukagakusho, the
Max Planck Society, and the Higher Education Funding Council for
England. The SDSS Web Site is http://www.sdss.org/.

The SDSS is managed by the Astrophysical Research Consortium for the
Participating Institutions. The Participating Institutions are the
American Museum of Natural History, Astrophysical Institute Potsdam,
University of Basel, University of Cambridge, Case Western Reserve
University, University of Chicago, Drexel University, Fermilab, the
Institute for Advanced Study, the Japan Participation Group, Johns
Hopkins University, the Joint Institute for Nuclear Astrophysics, the
Kavli Institute for Particle Astrophysics and Cosmology, the Korean
Scientist Group, the Chinese Academy of Sciences (LAMOST), Los Alamos
National Laboratory, the Max-Planck-Institute for Astronomy (MPIA),
the Max-Planck-Institute for Astrophysics (MPA), New Mexico State
University, Ohio State University, University of Pittsburgh,
University of Portsmouth, Princeton University, the United States
Naval Observatory, and the University of Washington.

This publication makes use of data products from the Two Micron All
Sky Survey, which is a joint project of the University of
Massachusetts and the Infrared Processing and Analysis
Center/California Institute of Technology, funded by the National
Aeronautics and Space Administration and the National Science
Foundation.

Finally, we have made use of the NASA/IPAC Extragalactic Database
(NED), which is operated by the Jet Propulsion Laboratory, California
Institute of Technology (Caltech) under contract with NASA. This
research has also made use of the VizieR catalogue access tool, CDS,
Strasbourg, France

{\it Facilities:} \facility{GALEX}, \facility{Sloan},
\facility{CTIO:1.5m}, \facility{KPNO:2.1m}, \facility{FLWO:2MASS},
\facility{CTIO:2MASS}, \facility{{\it Spitzer}}

\appendix
\section{Recalibrating the optical data}\label{recalib}
As mentioned in Section~\ref{opt_data}, a large subset of the original
SINGS optical images suffer from zero-point offsets whose origin has
not been fully identified. While some of the structural properties
presented in this paper, such as concentration indexes or
asymmetries, do not depend on these calibration issues, estimating
these offsets is essential if one wishes to employ these data in
stellar population studies.

The optical images were recalibrated using the catalog of aperture
photometry of Prugniel \& Heraudeau (1998). This compilation merges
photometric data from the literature obtained through different
methods: photoelectric photometry with diaphragms, simulated aperture
photometry from radial profiles and aperture measurements on
images.

Let $F_{\mathrm{ref,}\lambda}(r)$, $F_{\mathrm{ours,}\lambda}(r)$ and
$F_{\mathrm{true,}\lambda}(r)$ be the fluxes enclosed inside a
circular aperture of radius $r$ in the reference data, in our images
and in the actual galaxy, respectively. We can write the following:
\begin{eqnarray}
F_{\mathrm{true,}\lambda}(r) &=& C_{\lambda} F_{\mathrm{ours,}\lambda}(r)\label{eq_calib1}\\
F_{\mathrm{true,}\lambda}(r) &=& F_{\mathrm{ref,}\lambda}(r) - B_{\lambda}r^2\label{eq_calib2}\\
\frac{F_{\mathrm{ref,}\lambda}(r)}{r^2} &=& C_{\lambda} \frac{F_{\mathrm{ours,}\lambda}(r)}{r^2} + B_{\lambda}\label{eq_calib3}
\end{eqnarray}

In Eq.~\ref{eq_calib1} we are assuming that since we have masked most
relevant foreground and background objects, the shape of our radial
profiles is essentially correct, so we only need to multiply our data
by a certain recalibration factor $C_{\lambda}$ in order to get the
actual fluxes. As for the data from the literature, in
Eq.~\ref{eq_calib2} we suppose that they are not affected by
zero-point biases, but we allow for possible errors in the background
subtraction through the term $B_{\lambda}r^2$. While in our frames
there are always enough sky areas free from emission from the galaxy,
this might not be necessarily the case in some of the reference data
from the literature. Also, differences in the amount of background and
foreground objects that have been subtracted should also scale roughly
proportional to the aperture area, provided that these sources are
uniformly distributed throughout the field of view; to some extent the
$B_{\lambda}r^2$ term can also account for this effect, if
present. The sign of $B_{\lambda}$ can be either positive or negative
depending on whether the background in the reference data has been
over- or under-subtracted. Should the assumptions above be correct,
Eq.~\ref{eq_calib3} implies that the recalibration factor
$C_{\lambda}$ can be obtained from a simple linear fit.

We made use of the VizieR service (Ochsenbein et al$.$ 2000) to query
the catalog of Prugniel \& Heraudeau (1998) and retrieve all the
available aperture photometry for our galaxies. All magnitudes were
homogenized (in terms of filter systems) and converted to AB units
using the color transformations and AB magnitudes of Vega provided in
Fukugita et al$.$ (1995). We then used the IRAF task {\sc phot} to
perform aperture photometry on our images, using the same sets of
circular apertures as those compiled from the literature. The
recalibration factors were then obtained by comparing both sets of
data and applying a linear fit (Eq.~\ref{eq_calib3}).

The resulting values appear in Table~\ref{recalib_tab}. As an example,
Fig.~\ref{ngc1097_recalib} shows the results for NGC~1097 in the I
band. The small numbers next to each data-point show the radius of the
corresponding aperture in arcseconds. The behavior is clearly linear,
although very small apertures can sometimes depart from the general
trend, since they are prone to suffer from centering errors and seeing
effects. These points were excluded from the final fit, as well as
those showing different fluxes for exactly the same aperture
size. When there was only one available photometric measurement at a
given band, we assumed that $B_{\lambda}=0$ in Eq.~\ref{eq_calib3} and
computed the calibration factor $C_{\lambda}$ simply as the ratio of
the reference flux and ours.

\begin{figure}
\resizebox{!}{!}{\includegraphics{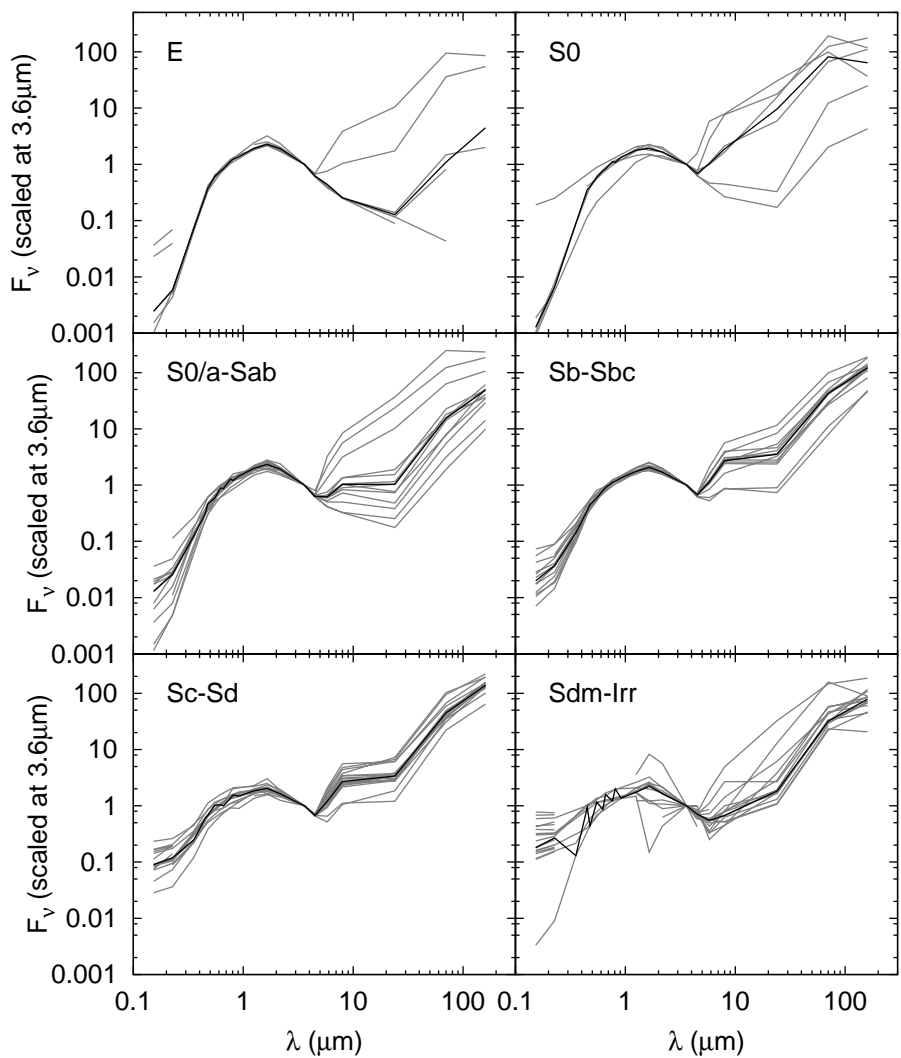}}
\caption{Multi-wavelength asymptotic magnitudes of the SINGS galaxies,
sorted out into different morphological types. The black solid line
indicates the median SEDs in each panel. Small differences between
intercalated Johnson-Cousins and Sloan bands can lead to the observed
saw-tooth shape in late-type spirals. Note that optical data are
missing for some galaxies, when they could not be recalibrated and
Sloan data were not available.\label{seds_norm}}
\end{figure}

\begin{figure}
\resizebox{!}{!}{\includegraphics{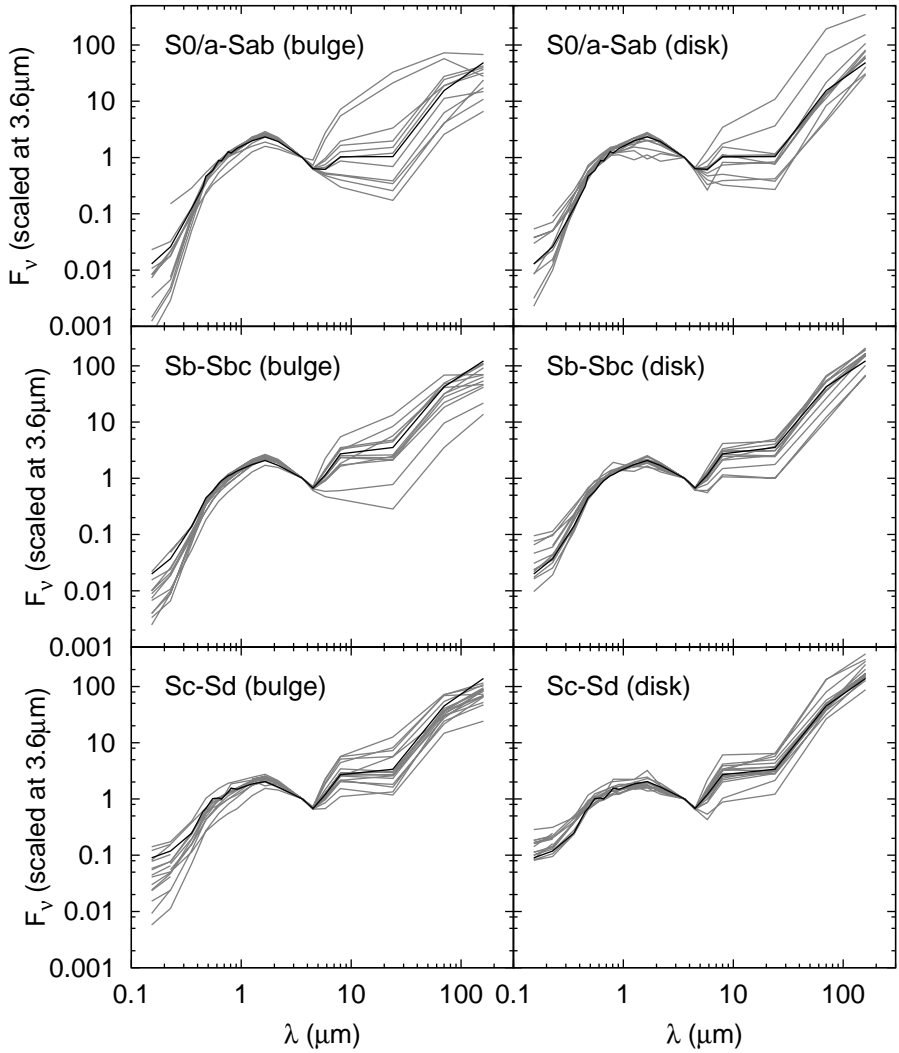}}
\caption{Spectral energy distributions of the bulge- and
disk-dominated regions of spiral galaxies galaxies. As a reference,
the black solid line shows the median SED of galaxies of the
corresponding morphological type, taken as a whole (the same curves as
those in Fig.~\ref{seds_norm}). \label{seds_norm_bd}}
\end{figure}

\begin{figure}
\resizebox{!}{0.7\vsize}{\includegraphics{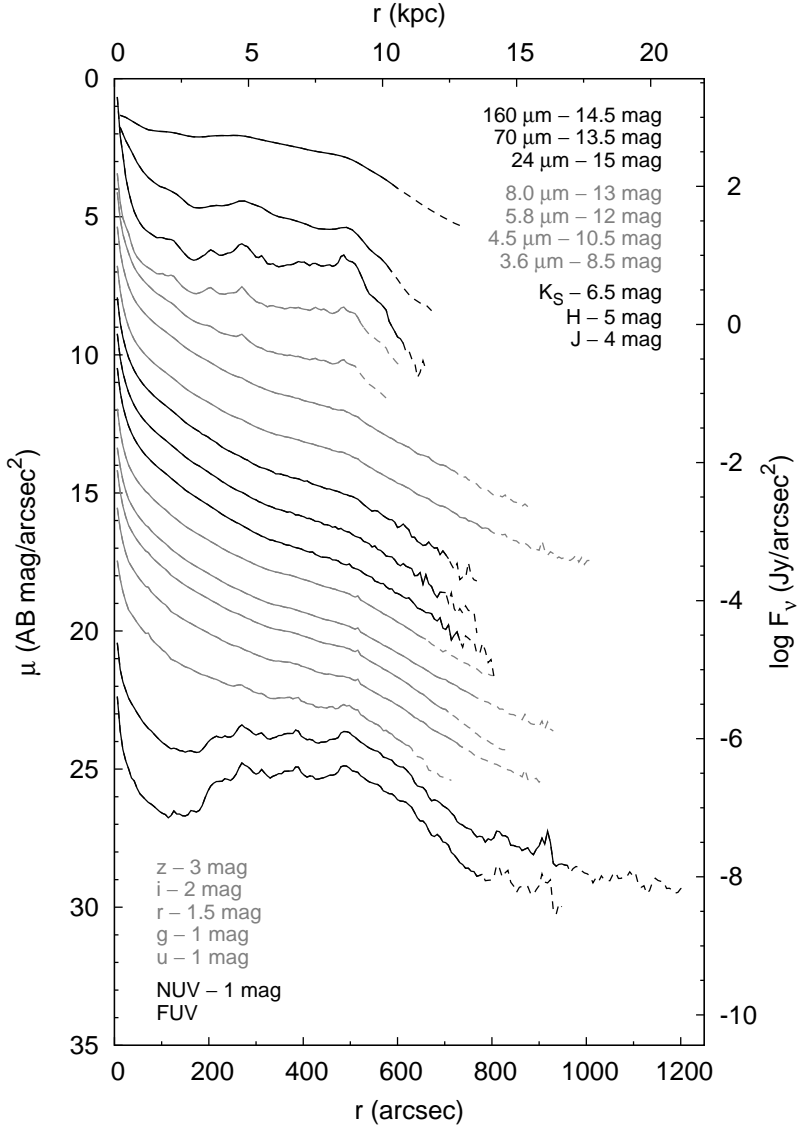}}
\caption{Multi-wavelength surface brightness profiles for NGC~3031
(see the on-line edition for the full version of this figure including
all galaxies in the sample). The profiles have been shifted for
displaying purposes; the corresponding offsets in magnitudes are
quoted next to each label. The profiles are arranged in order of
decreasing wavelength, from top to bottom, as shown by the labels. For
the sake of clarity, black and gray lines are used to group both the
profiles and their labels according to their wavelength range (GALEX,
optical, 2MASS, IRAC and MIPS, respectively). Errorbars are not shown
for clarity (see Tables~\ref{uv_opt_2mass_tab},
\ref{uv_sdss_2mass_tab} and \ref{irac_mips_tab}). The solid profiles
have been truncated when $\Delta \mu > 0.3$\,mag, and then continue
with dashed lines until $\Delta \mu > 1$\,mag. Note that the large
uncertainties in the outermost spatial regions (i.e$.$ those marked
with dashed lines) are mostly due to large-scale errors in the
background estimation, but do not necessarily imply
non-detections. Emission from the galaxy can be clearly seen in these
regions above the local noise, although large-scale background
variations preclude a more reliable determination of the
azimuthally-averaged flux density along these outer isophotes.
\label{ngc3031_allprofs_offset}}
\end{figure}

\begin{figure}
\resizebox{!}{0.7\vsize}{\includegraphics{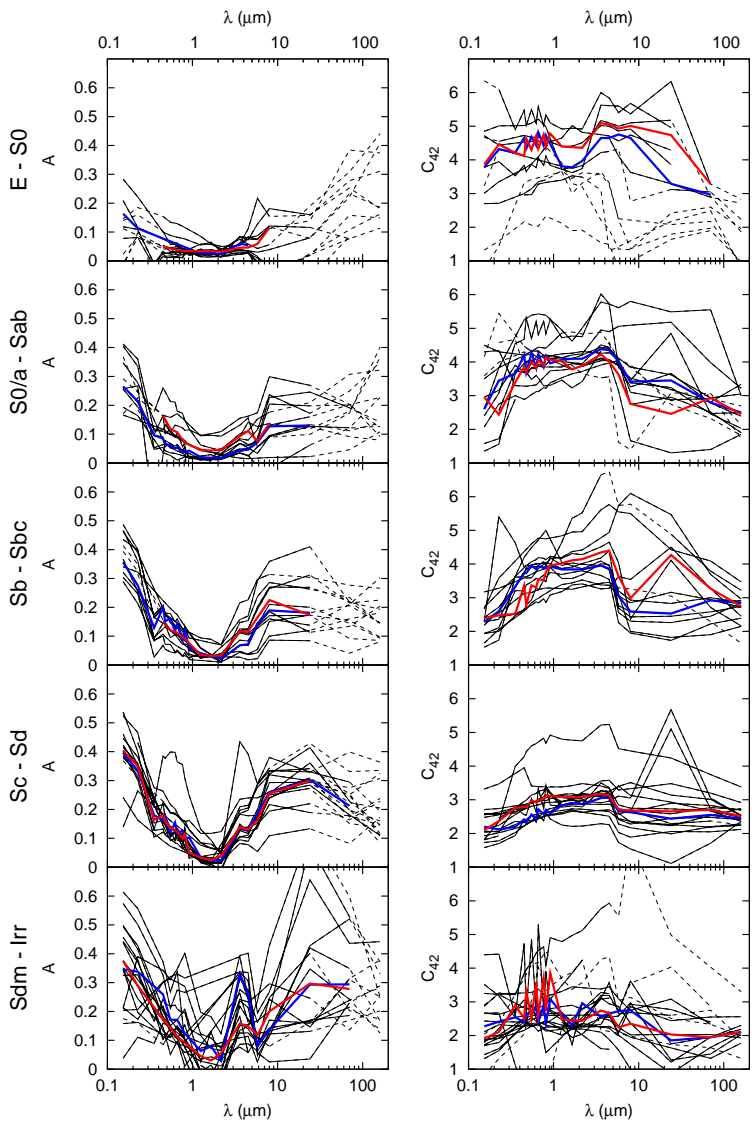}}
\caption{Left: Asymmetry of the SINGS galaxies as a function of
wavelength, in bins of morphological type. Dashed lines correspond to
those bands in which we are unable to resolve structures smaller than
0.5\,kpc at the particular distance of each galaxy (note that the
meaning of dashed lines is different in the right column, see
below). The red (blue) lines show the median values for the galaxies
further (closer) than the median distance within each morphological
bin (see Section~\ref{morpho_lambda}). Dashed lines were not used when
computing the red and blue median curves. The galaxy with large
optical and infrared asymmetries in the Sc-Sd panel is NGC~5474 (see
text). Note that the larger dispersion seen in the Sdm-Irr panel
(especially at 3.6\,$\micron$ and 4.5\,$\micron$) is probably due to
faint unremoved background or foreground sources, since these galaxies
have lower surface brightness. Right: Concentration index at different
wavelengths. Dashed lines are used when the inner radius $r_{20}$ used
to compute $C_{42}$ is smaller than the innermost point of our
profiles. These values are then just lower limits. Red and blue lines
have the same meaning as in the asymmetry panels.\label{conc_asym}}
\end{figure}

\begin{figure}
\resizebox{!}{0.7\vsize}{\includegraphics{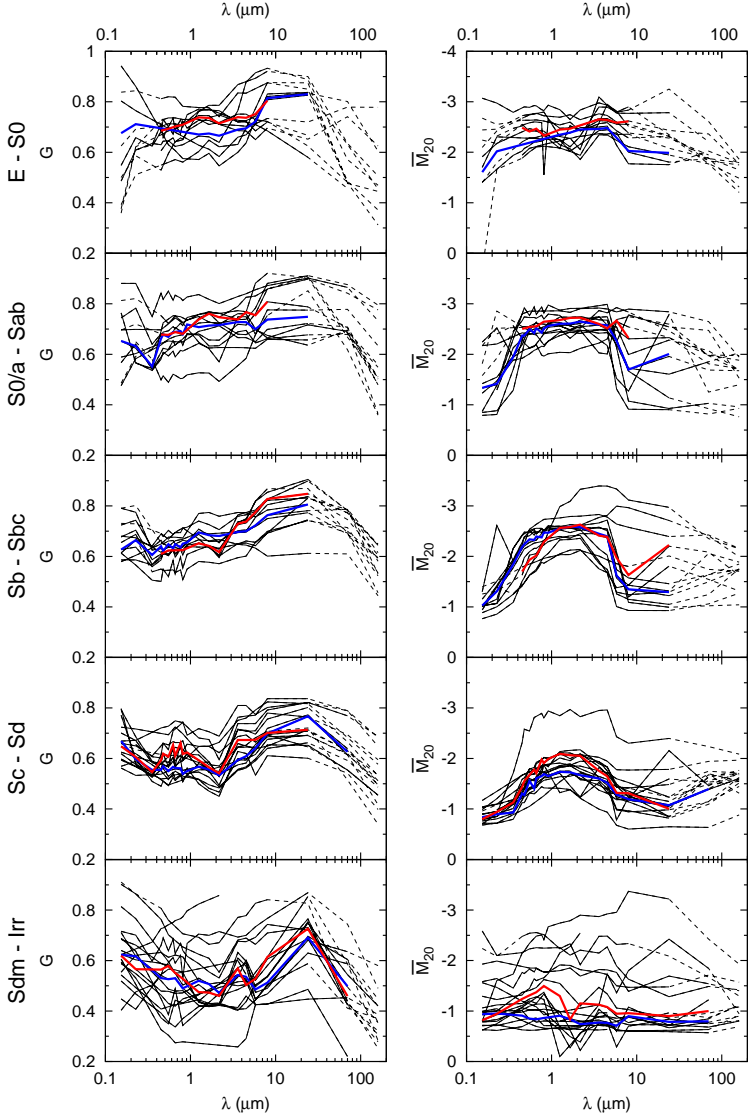}}
\caption{Gini coefficient (left) and normalized second-order moment of
the brightest 20\% of the emission (right) as a function of wavelength
and in bins of morphological type. Dashed lines are used when the FWHM
in a given band at the particular distance of each galaxy is larger
than 0.5\,kpc. The red (blue) lines show the median values of the
galaxies further (closer) than the median distance within each
morphological bin (see Section~\ref{morpho_lambda}). Dashed lines were
not used when computing these median curves.\label{gini_m20}}
\end{figure}

\begin{figure}
\resizebox{!}{!}{\includegraphics{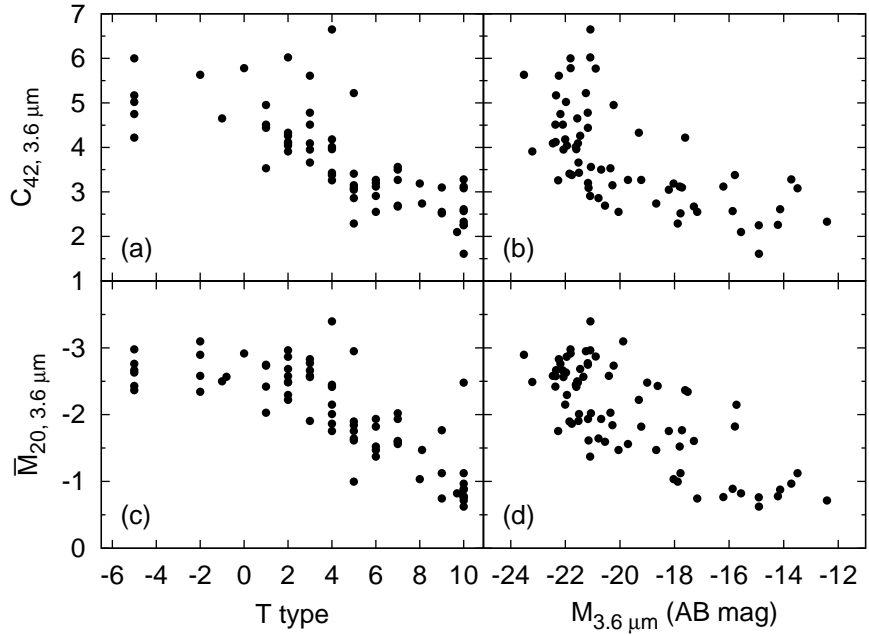}}
\caption{Top row: concentration index of the SINGS galaxies at
3.6\,$\micron$ as a function of their Hubble type (a) and their
absolute magnitude at 3.6\,$\micron$ (b). Bottom row: normalized
second-order moment of the brightest 20\% of the emission as a
function of the same quantities.
\label{cindex_mag}}
\end{figure}

The statistical uncertainty of the recalibration factors yielded by
the fitting procedure is typically less than 5\%. This is the error of
the recalibrated data relative to the original values found in the
literature, which of course carry their own uncertainties. The latter
are difficult to constrain, as they are not quoted in the catalog of
Prugniel \& Heraudeau (1998). Nevertheless, we reckon that a final
zero-point error of 10\%-15\% should be adopted when using the
recalibrated optical data. In Fig.~\ref{seds_norm} galaxies with SDSS
data and those with recalibrated optical data are plotted together;
the small dispersion constitutes a sanity check for our recalibration
procedure. Note that recalibration factors are not provided for some
SINGS galaxies (mainly dwarfs) that do appear in the catalog of
Prugniel \& Heraudeau (1998), since the resulting optical data seemed
to be largely discrepant from the overall SED.

\clearpage
\begin{deluxetable}{lrrrrrcrrccr}
\tabletypesize{\scriptsize}
\tablecolumns{12}
\setlength{\tabcolsep}{0.045in}
\tablecaption{Sample\label{sample_tab}}
\tablewidth{0pt}
\tablehead{
\colhead{Object name} & \colhead{RA$_{2000}$} & \colhead{DEC$_{2000}$} & \colhead{$2a$} & \colhead{$2b$} &\colhead{P.A.} & \colhead{E(B$-$V)} & \colhead{dist} & \colhead{$T$} & \colhead{Morphological} & \colhead{Optical} & \colhead{$r_P$}\\
\colhead{} & \colhead{(h:m:s)} & \colhead{(d:m:s)} & \colhead{(arcmin)} & \colhead{(arcmin)} & \colhead{(deg)} & \colhead{(mag)} & \colhead{(Mpc)} & \colhead{type} & \colhead{type} & \colhead{data} & \colhead{(arcsec)}\\
\colhead{(1)} & \colhead{(2)} & \colhead{(3)} & \colhead{(4)} & \colhead{(5)} & \colhead{(6)} & \colhead{(7)} & \colhead{(8)} & \colhead{(9)} & \colhead{(10)} & \colhead{(11)} & \colhead{(12)}}
\startdata
   NGC~0024 & 00 09 56.5 & $-$24 57 47.3 &   5.8 &   1.3 & 46 & 0.020 & 8.2 & 5 & SA(s)c & recal &   152\\
   NGC~0337 & 00 59 50.1 & $-$07 34 40.7 &   2.9 &   1.8 & 310 & 0.112 &  25 & 7 & SB(s)d & \nodata &    63\\
   NGC~0584 & 01 31 20.7 & $-$06 52 05.0 &   4.2 &   2.3 & 55 & 0.042 &  28 & $-$5 & E4 & recal &    47\\
   NGC~0628 & 01 36 41.8 & 15 47 00.5 &  10.5 &   9.5 & 25 & 0.070 &  11 & 5 & SA(s)c & \nodata &   239\\
   NGC~0855 & 02 14 03.6 & 27 52 37.8 &   2.6 &   1.0 & 60 & 0.072 & 9.7 & $-$5 & E & \nodata &    50\\
   NGC~0925 & 02 27 16.9 & 33 34 45.0 &  10.5 &   5.9 & 282 & 0.076 & 9.3 & 7 & SAB(s)d & recal &   280\\
   NGC~1097 & 02 46 19.1 & $-$30 16 29.7 &   9.3 &   6.3 & 310 & 0.027 &  15 & 3 & SB(s)b & recal &   232\\
   NGC~1266 & 03 16 00.7 & $-$02 25 38.5 &   1.5 &   1.0 & 290 & 0.098 &  31 & $-$2 & (R')SB0(rs) pec & \nodata &    34\\
\enddata
\tablecomments{Sample. (1): Galaxy name. (2): RA(J2000) of the galaxy center. (3): DEC(J2000) of the galaxy center. (4),(5): Apparent major and minor isophotal diameters at $\mu_{B}$=25 mag arcsec$^{-2}$ from the RC3 catalog. (6): Position angle from RC3. \dag The PA and axis ratio for NGC~5194 differ from those in the RC3, which are affected by the presence of NGC~5195. (7): Galactic color excess from Schlegel et al$.$ (1998). (8): Distance to the galaxy, rounded to the nearest Mpc when larger than 10 Mpc, taken from Gil de Paz et al$.$ (2007) and Kennicutt et al$.$ (2003). (9): Morphological type $T$ as given in the RC3 catalog.  (10): Full description of the morphological type from the RC3. (11): Optical data used for each galaxy: from the Sloan Digital Sky Survey (SDSS) or recalibrated with compiled aperture photometry (recal, see Section~\ref{opt_data} and Appendix~\ref{recalib}). Galaxies lacking any label in this column have optical images with unreliable zero-points. Neither surface-brightness profiles nor asymptotic magnitudes are provided for these galaxies in the optical bands. However, we do provide morphological estimators, since they are not affected by zero-point errors. (12): Petrosian radius at $\eta=0.2$, measured on the 3.6\,$\micron$ images (except for a few galaxies where no meaningful solution was found, and the $B$-band images were used instead).}
\end{deluxetable}

\clearpage
\begin{deluxetable}{rrrrrrrrrrr}
\tabletypesize{\scriptsize}
\tablecolumns{11}
\tablecaption{UV, optical and nIR surface photometry.\label{uv_opt_2mass_tab}}
\tablewidth{0pt}
\rotate
\tablehead{
\colhead{r} & \colhead{r} & \colhead{FUV} & \colhead{NUV} & \colhead{$B$} & \colhead{$V$} & \colhead{$R$} & \colhead{$I$} & \colhead{$J$} & \colhead{$H$} & \colhead{$K_S$}\\
\colhead{} & \colhead{} & \colhead{0.153\,$\micron$} & \colhead{0.227\,$\micron$} & \colhead{0.45\,$\micron$} & \colhead{0.55\,$\micron$} & \colhead{0.66\,$\micron$} & \colhead{0.81\,$\micron$} & \colhead{1.25\,$\micron$} & \colhead{1.65\,$\micron$} & \colhead{2.17\,$\micron$}\\
\colhead{(arcsec)} & \colhead{(kpc)} & \colhead{(mag/\sq)} & \colhead{(mag/\sq)} & \colhead{(mag/\sq)} & \colhead{(mag/\sq)} & \colhead{(mag/\sq)} & \colhead{(mag/\sq)} & \colhead{(mag/\sq)} & \colhead{(mag/\sq)} & \colhead{(mag/\sq)}}
\startdata
\cutinhead{NGC~7331}
    6 &   0.4 & 	25.03$\pm$0.05  & 22.87$\pm$0.01  & 	17.69$\pm$0.01  & 16.82$\pm$0.01  & 16.21$\pm$0.01  & 15.92$\pm$0.01  & 	15.44$\pm$0.01  & 15.19$\pm$0.01  & 15.38$\pm$0.01   \\
   12 &   0.9 & 	25.21$\pm$0.06  & 23.15$\pm$0.01  & 	18.54$\pm$0.01  & 17.75$\pm$0.01  & 17.17$\pm$0.01  & 16.87$\pm$0.01  & 	16.28$\pm$0.01  & 16.03$\pm$0.01  & 16.24$\pm$0.01   \\
   18 &   1.3 & 	24.75$\pm$0.04  & 23.07$\pm$0.01  & 	19.06$\pm$0.01  & 18.32$\pm$0.01  & 17.74$\pm$0.01  & 17.45$\pm$0.01  & 	16.81$\pm$0.01  & 16.56$\pm$0.01  & 16.77$\pm$0.01   \\
   24 &   1.7 & 	24.32$\pm$0.03  & 22.91$\pm$0.01  & 	19.41$\pm$0.01  & 18.72$\pm$0.01  & 18.15$\pm$0.01  & 17.87$\pm$0.01  & 	17.16$\pm$0.01  & 16.91$\pm$0.01  & 17.10$\pm$0.01   \\
   30 &   2.2 & 	24.34$\pm$0.02  & 22.99$\pm$0.01  & 	19.67$\pm$0.01  & 19.05$\pm$0.01  & 18.50$\pm$0.01  & 18.23$\pm$0.01  & 	17.46$\pm$0.01  & 17.18$\pm$0.01  & 17.34$\pm$0.01   \\
   36 &   2.6 & 	24.68$\pm$0.02  & 23.33$\pm$0.01  & 	19.99$\pm$0.01  & 19.36$\pm$0.01  & 18.81$\pm$0.01  & 18.52$\pm$0.01  & 	17.77$\pm$0.01  & 17.45$\pm$0.01  & 17.59$\pm$0.01   \\
\enddata
\tablecomments{Surface photometry in the UV, optical and near-IR passbands for those SINGS galaxies lacking SDSS data. Only one sample galaxy is shown in the printed version, check the online edition for the full table. Radius is measured along the semimajor axis. The surface brightness is expressed in AB mag/arcsec$^2$. All values are corrected for foreground galactic extinction. The quoted uncertainties include photometric and sky errors, but not zero-point errors (see Section~\ref{surf_phot} for details). The optical values (columns 5 through 8) have been recalibrated to correct zero-point offsets present in the original data (see Section~\ref{recalib}). Again, the errors quoted here do not include the uncertainty resulting from the recalibration procedure.}
\end{deluxetable}

\clearpage
\begin{deluxetable}{rrrrrrrrrrrr}
\tabletypesize{\scriptsize}
\tablecolumns{12}
\tablecaption{UV, optical (SDSS) and nIR surface photometry\label{uv_sdss_2mass_tab}}
\tablewidth{0pt}
\rotate
\tablehead{
\colhead{r} & \colhead{r} & \colhead{FUV} & \colhead{NUV} & \colhead{$u$} & \colhead{$g$} & \colhead{$r$} & \colhead{$i$} & \colhead{$z$} & \colhead{$J$} & \colhead{$H$} & \colhead{$K_S$}\\
\colhead{} & \colhead{} & \colhead{0.153\,$\micron$} & \colhead{0.227\,$\micron$} & \colhead{0.354\,$\micron$} & \colhead{0.477\,$\micron$} & \colhead{0.623\,$\micron$} & \colhead{0.762\,$\micron$} & \colhead{0.913\,$\micron$} & \colhead{1.25\,$\micron$} & \colhead{1.65\,$\micron$} & \colhead{2.17\,$\micron$}\\
\colhead{(arcsec)} & \colhead{(kpc)} & \colhead{(mag/\sq)} & \colhead{(mag/\sq)} & \colhead{(mag/\sq)} & \colhead{(mag/\sq)} & \colhead{(mag/\sq)} & \colhead{(mag/\sq)} & \colhead{(mag/\sq)} & \colhead{(mag/\sq)} & \colhead{(mag/\sq)} & \colhead{(mag/\sq)}}
\startdata
\cutinhead{NGC~3184}
    6 &   0.3 & 	24.15$\pm$0.12  & 23.25$\pm$0.05  & 	22.03$\pm$0.02  & 20.55$\pm$0.01  & 19.81$\pm$0.01  & 19.41$\pm$0.01  & 19.17$\pm$0.01  & 	18.79$\pm$0.01  & 18.58$\pm$0.01  & 18.83$\pm$0.02   \\
   12 &   0.5 & 	25.52$\pm$0.21  & 24.85$\pm$0.10  & 	22.73$\pm$0.03  & 21.14$\pm$0.01  & 20.40$\pm$0.01  & 19.99$\pm$0.01  & 19.80$\pm$0.01  & 	19.44$\pm$0.01  & 19.29$\pm$0.02  & 19.55$\pm$0.02   \\
   18 &   0.8 & 	25.57$\pm$0.12  & 24.77$\pm$0.05  & 	22.92$\pm$0.04  & 21.39$\pm$0.01  & 20.69$\pm$0.01  & 20.30$\pm$0.01  & 20.12$\pm$0.01  & 	19.78$\pm$0.01  & 19.60$\pm$0.02  & 19.85$\pm$0.02   \\
   24 &   1.0 & 	24.83$\pm$0.07  & 24.16$\pm$0.03  & 	22.83$\pm$0.04  & 21.48$\pm$0.01  & 20.85$\pm$0.01  & 20.49$\pm$0.01  & 20.32$\pm$0.01  & 	19.97$\pm$0.01  & 19.81$\pm$0.02  & 20.04$\pm$0.02   \\
   30 &   1.3 & 	24.38$\pm$0.05  & 23.97$\pm$0.03  & 	22.84$\pm$0.04  & 21.61$\pm$0.01  & 21.03$\pm$0.01  & 20.69$\pm$0.01  & 20.53$\pm$0.02  & 	20.20$\pm$0.02  & 20.00$\pm$0.02  & 20.29$\pm$0.03   \\
   36 &   1.5 & 	24.70$\pm$0.05  & 24.22$\pm$0.03  & 	23.16$\pm$0.05  & 21.89$\pm$0.01  & 21.29$\pm$0.01  & 20.95$\pm$0.01  & 20.79$\pm$0.02  & 	20.48$\pm$0.02  & 20.30$\pm$0.02  & 20.58$\pm$0.03   \\
\enddata
\tablecomments{Surface photometry in the UV, optical and near-IR passbands for those SINGS galaxies having SDSS data. Only one sample galaxy is shown in the printed version, check the online edition for the full table. The radius is measured along the semimajor axis. The surface brightness is in AB (mag/arcsec$^2$). All values are corrected for foreground galactic extinction. The quoted uncertainties include photometric and sky errors, but not zero-point errors (see Section~\ref{surf_phot} for details).}
\end{deluxetable}

\clearpage
\begin{deluxetable}{rrrrrrrrr}
\tabletypesize{\scriptsize}
\tablecolumns{9}
\tablecaption{IRAC and MIPS surface photometry\label{irac_mips_tab}}
\tablewidth{0pt}
\tablehead{
\colhead{r} & \colhead{r} & \colhead{$\log I_{3.6\micron}$} & \colhead{$\log I_{4.5\micron}$} & \colhead{$\log I_{5.8\micron}$} & \colhead{$\log I_{8.0\micron}$} & \colhead{$\log I_{24\micron}$} & \colhead{$\log I_{70\micron}$} & \colhead{$\log I_{160\micron}$}\\
\colhead{(arcsec)} & \colhead{(kpc)} & \colhead{(Jy/\sq)} & \colhead{(Jy/\sq)} & \colhead{(Jy/\sq)} & \colhead{(Jy/\sq)} & \colhead{(Jy/\sq)} & \colhead{(Jy/\sq)} & \colhead{(Jy/\sq)}}
\startdata
\cutinhead{NGC~3184}
    6 &   0.3 & 	-4.28$\pm$0.01  & -4.46$\pm$0.01  & -4.29$\pm$0.01  & -3.97$\pm$0.01  & 	-3.60$\pm$0.01  & 	                  &                    \\
   12 &   0.5 & 	-4.58$\pm$0.01  & -4.77$\pm$0.01  & -4.74$\pm$0.01  & -4.54$\pm$0.01  & 	-4.14$\pm$0.01  & 	-3.06$\pm$0.01  & -2.75$\pm$0.01   \\
   18 &   0.8 & 	-4.72$\pm$0.01  & -4.90$\pm$0.01  & -4.81$\pm$0.01  & -4.57$\pm$0.01  & 	-4.52$\pm$0.01  & 	                  &                    \\
   24 &   1.0 & 	-4.77$\pm$0.01  & -4.96$\pm$0.01  & -4.77$\pm$0.01  & -4.45$\pm$0.01  & 	-4.49$\pm$0.01  & 	-3.39$\pm$0.02  & -2.82$\pm$0.01   \\
   30 &   1.3 & 	-4.86$\pm$0.01  & -5.04$\pm$0.01  & -4.83$\pm$0.01  & -4.49$\pm$0.01  & 	-4.52$\pm$0.01  & 	                  &                    \\
   36 &   1.5 & 	-4.96$\pm$0.01  & -5.15$\pm$0.01  & -4.94$\pm$0.01  & -4.61$\pm$0.01  & 	-4.62$\pm$0.01  & 	-3.47$\pm$0.02  & -2.88$\pm$0.01   \\
\enddata
\tablecomments{Surface photometry in the IRAC and MIPS bands. Check the online edition for the full table with all galaxies. The radius is measured along the semimajor axis. The surface brightness is in Jy/arcsec$^2$. All values are corrected for foreground galactic extinction. The quoted uncertainties include photometric and sky errors, but not zero-point errors (see Section~\ref{surf_phot} for details).}
\end{deluxetable}

\clearpage
\begin{deluxetable}{rrrrrrrrrr}
\tabletypesize{\scriptsize}
\tablecolumns{10}
\tablecaption{UV, optical and nIR asymptotic magnitudes.\label{asmag_uv_opt_2mass_tab}}
\tablewidth{0pt}
\rotate
\tablehead{
\colhead{Galaxy} & \colhead{FUV} & \colhead{NUV} & \colhead{$B$} & \colhead{$V$} & \colhead{$R$} & \colhead{$I$} & \colhead{$J$} & \colhead{$H$} & \colhead{$K_S$}\\
\colhead{} & \colhead{0.153\,$\micron$} & \colhead{0.227\,$\micron$} & \colhead{0.45\,$\micron$} & \colhead{0.55\,$\micron$} & \colhead{0.66\,$\micron$} & \colhead{0.81\,$\micron$} & \colhead{1.25\,$\micron$} & \colhead{1.65\,$\micron$} & \colhead{2.17\,$\micron$}\\
\colhead{} & \colhead{(mag)} & \colhead{(mag)} & \colhead{(mag)} & \colhead{(mag)} & \colhead{(mag)} & \colhead{(mag)} & \colhead{(mag)} & \colhead{(mag)} & \colhead{(mag)}}
\startdata
\cutinhead{Galaxies with recalibrated optical data}
        NGC~0024 & 14.039$\pm$0.004  & 13.757$\pm$0.003  & 11.912$\pm$0.002  & 11.341$\pm$0.002  & 11.276$\pm$0.002  & 10.553$\pm$0.002  & 10.501$\pm$0.006  & 10.371$\pm$0.011  & 10.714$\pm$0.010 \\
        NGC~0584 & 17.506$\pm$0.059  & 15.631$\pm$0.015  & 11.506$\pm$0.003  & 10.491$\pm$0.003  & 10.194$\pm$0.004  &  9.759$\pm$0.004  &  9.065$\pm$0.005  &  8.758$\pm$0.009  &  9.099$\pm$0.009 \\
        NGC~0925 & 12.072$\pm$0.001  & 11.852$\pm$0.001  & 10.471$\pm$0.003  &  9.972$\pm$0.003  & \nodata &  9.648$\pm$0.006  &  9.464$\pm$0.002  &  9.310$\pm$0.007  &  9.604$\pm$0.005 \\
        NGC~1097 & 12.542$\pm$0.004  & 12.167$\pm$0.003  &  9.860$\pm$0.001  &  9.300$\pm$0.001  &  8.958$\pm$0.001  &  8.623$\pm$0.001  &  8.019$\pm$0.001  &  7.825$\pm$0.002  &  8.050$\pm$0.003 \\
        NGC~1291 & 14.229$\pm$0.015  & 13.380$\pm$0.005  &  9.439$\pm$0.001  &  8.675$\pm$0.002  &  8.282$\pm$0.003  &  7.925$\pm$0.005  &  7.374$\pm$0.001  &  7.305$\pm$0.003  &  7.444$\pm$0.001 \\
        NGC~1316 & 15.071$\pm$0.056  & 13.079$\pm$0.026  &  9.080$\pm$0.014  &  8.353$\pm$0.008  &  7.968$\pm$0.010  &  7.709$\pm$0.005  &  7.438$\pm$0.001  &  7.373$\pm$0.001  &  7.547$\pm$0.001 \\
        NGC~1377 & \nodata & \nodata & 13.041$\pm$0.010  & 12.679$\pm$0.008  & 12.326$\pm$0.006  & 12.007$\pm$0.008  & 11.468$\pm$0.004  & 11.274$\pm$0.015  & 11.515$\pm$0.013 \\
        NGC~1404 & 16.413$\pm$0.008  & 15.275$\pm$0.013  & 10.763$\pm$0.011  &  9.988$\pm$0.010  &  9.598$\pm$0.011  &  9.175$\pm$0.010  &  8.648$\pm$0.001  &  8.447$\pm$0.003  &  8.644$\pm$0.002 \\
\enddata
\tablecomments{Asymptotic magnitudes in the UV, optical and near-IR passbands for those SINGS galaxies lacking SDSS data. All values are in AB magnitudes, and corrected for foreground galactic extinction. The quoted uncertainties are just the statitistical errors resulting from the fit applied to the growth curve, but do not include zero-point errors (see Section~\ref{asmag} for details). The optical values (columns 5 through 8) have been recalibrated to correct zero-point offsets present in the original data (see Section~\ref{recalib}). Again, the errors quoted here do not include the uncertainty resulting from the recalibration procedure.}
\end{deluxetable}

\clearpage
\begin{deluxetable}{rrrrrrrrrrr}
\tabletypesize{\scriptsize}
\tablecolumns{11}
\tablecaption{UV, optical (SDSS) and nIR asymptotic magnitudes.\label{asmag_uv_sdss_2mass_tab}}
\tablewidth{0pt}
\rotate
\tablehead{
\colhead{Galaxy} & \colhead{FUV} & \colhead{NUV} & \colhead{$u$} & \colhead{$g$} & \colhead{$r$} & \colhead{$i$} & \colhead{$z$} & \colhead{$J$} & \colhead{$H$} & \colhead{$K_S$}\\
\colhead{} & \colhead{0.153\,$\micron$} & \colhead{0.227\,$\micron$} & \colhead{0.354\,$\micron$} & \colhead{0.477\,$\micron$} & \colhead{0.623\,$\micron$} & \colhead{0.762\,$\micron$} & \colhead{0.913\,$\micron$} & \colhead{1.25\,$\micron$} & \colhead{1.65\,$\micron$} & \colhead{2.17\,$\micron$}\\
\colhead{} & \colhead{(mag)} & \colhead{(mag)} & \colhead{(mag)} & \colhead{(mag)} & \colhead{(mag)} & \colhead{(mag)} & \colhead{(mag)} & \colhead{(mag)} & \colhead{(mag)} & \colhead{(mag)}}
\startdata
        NGC~2841 & 13.527$\pm$0.019  & 13.014$\pm$0.010  & 11.229$\pm$0.003  &  9.696$\pm$0.001  &  8.953$\pm$0.001  &  8.515$\pm$0.001  &  8.269$\pm$0.001  &  7.910$\pm$0.001  &  7.687$\pm$0.001  &  7.934$\pm$0.001 \\
        NGC~2976 & 13.227$\pm$0.003  & 12.707$\pm$0.003  & 11.464$\pm$0.002  & 10.293$\pm$0.001  &  9.749$\pm$0.002  &  9.471$\pm$0.001  &  9.293$\pm$0.002  &  9.163$\pm$0.002  &  9.046$\pm$0.002  &  9.312$\pm$0.003 \\
        NGC~3049 & \nodata & 14.736$\pm$0.004  & 13.787$\pm$0.007  & 12.884$\pm$0.008  & 12.350$\pm$0.010  & 12.112$\pm$0.011  & 11.909$\pm$0.018  & 11.621$\pm$0.007  & 11.631$\pm$0.009  & 11.731$\pm$0.022 \\
        NGC~3031 & 10.744$\pm$0.002  & 10.344$\pm$0.002  &  8.640$\pm$0.004  &  7.177$\pm$0.002  &  6.487$\pm$0.001  &  6.068$\pm$0.001  &  5.861$\pm$0.001  &  5.601$\pm$0.001  &  5.411$\pm$0.001  &  5.632$\pm$0.001 \\
        NGC~3034 & 12.345$\pm$0.016  & 11.313$\pm$0.005  &  9.669$\pm$0.001  &  8.234$\pm$0.001  &  7.542$\pm$0.001  &  7.231$\pm$0.001  &  7.030$\pm$0.001  &  6.593$\pm$0.001  &  6.336$\pm$0.001  &  6.433$\pm$0.001 \\
        NGC~3190 & 17.372$\pm$0.012  & 15.803$\pm$0.008  & 12.842$\pm$0.008  & 11.317$\pm$0.002  & 10.545$\pm$0.001  & 10.136$\pm$0.001  &  9.885$\pm$0.002  &  9.360$\pm$0.001  &  9.110$\pm$0.002  &  9.272$\pm$0.001 \\
        NGC~3184 & 12.476$\pm$0.002  & 12.177$\pm$0.001  & 11.236$\pm$0.001  & 10.078$\pm$0.001  &  9.582$\pm$0.001  &  9.298$\pm$0.001  &  9.186$\pm$0.002  &  9.019$\pm$0.004  &  8.817$\pm$0.005  &  9.094$\pm$0.004 \\
        NGC~3198 & 12.985$\pm$0.002  & 12.777$\pm$0.001  & 11.600$\pm$0.017  & 10.714$\pm$0.006  & 10.283$\pm$0.004  & 10.001$\pm$0.004  &  9.808$\pm$0.006  &  9.584$\pm$0.004  &  9.397$\pm$0.006  &  9.603$\pm$0.004 \\
\enddata
\tablecomments{Asymptotic magnitudes in the UV, optical and near-IR passbands for those SINGS galaxies with available SDSS data. All values are in AB magnitudes, and are corrected for foreground galactic extinction. The quoted uncertainties are just the statitistical errors resulting from the fit applied to the growth curve, but do not include zero-point errors (see Section~\ref{asmag} for details).}
\end{deluxetable}

\clearpage
\begin{deluxetable}{rrrrrrrr}
\tabletypesize{\scriptsize}
\tablecolumns{8}
\tablecaption{IRAC and MIPS asymptotic magnitudes.\label{asmag_irac_mips_tab}}
\tablewidth{0pt}
\tablehead{
\colhead{Galaxy} & \colhead{$\log F_{3.6\micron}$} & \colhead{$\log F_{4.5\micron}$} & \colhead{$\log F_{5.8\micron}$} & \colhead{$\log F_{8.0\micron}$} & \colhead{$\log F_{24\micron}$} & \colhead{$\log F_{70\micron}$} & \colhead{$\log F_{160\micron}$}\\
\colhead{} & \colhead{(Jy)} & \colhead{(Jy)} & \colhead{(Jy)} & \colhead{(Jy)} & \colhead{(Jy)} & \colhead{(Jy)} & \colhead{(Jy)}}
\startdata
        NGC~0024 & -0.984$\pm$0.001  & -1.153$\pm$0.002  & -1.169$\pm$0.002  & -0.936$\pm$0.001  & -0.904$\pm$0.004  &  0.362$\pm$0.005  &  0.825$\pm$0.024 \\
        NGC~0337 & -1.006$\pm$0.001  & -1.164$\pm$0.002  & -0.810$\pm$0.001  & -0.430$\pm$0.001  & -0.146$\pm$0.004  &  1.006$\pm$0.002  &  1.279$\pm$0.009 \\
        NGC~0584 & -0.450$\pm$0.001  & -0.661$\pm$0.001  & -0.850$\pm$0.001  & -1.051$\pm$0.001  & -1.380$\pm$0.008  & -0.538$\pm$0.060  & \nodata\\
        NGC~0628 & -0.076$\pm$0.002  & -0.241$\pm$0.002  &  0.046$\pm$0.002  &  0.461$\pm$0.002  &  0.498$\pm$0.003  &  1.536$\pm$0.002  &  2.087$\pm$0.001 \\
        NGC~0855 & -1.371$\pm$0.003  & -1.544$\pm$0.002  & -1.493$\pm$0.004  & -1.356$\pm$0.002  & -1.132$\pm$0.006  &  0.186$\pm$0.009  &  0.366$\pm$0.015 \\
        NGC~0925 & -0.494$\pm$0.002  & -0.656$\pm$0.002  & -0.508$\pm$0.003  & -0.158$\pm$0.003  & -0.046$\pm$0.007  &  1.158$\pm$0.005  &  1.621$\pm$0.002 \\
        NGC~1097 &  0.089$\pm$0.001  & -0.084$\pm$0.001  &  0.147$\pm$0.006  &  0.504$\pm$0.006  &  0.816$\pm$0.002  &  1.737$\pm$0.002  &  2.165$\pm$0.001 \\
        NGC~1266 & -1.269$\pm$0.002  & -1.377$\pm$0.002  & -1.271$\pm$0.011  & -1.053$\pm$0.008  & -0.074$\pm$0.004  &  1.013$\pm$0.005  &  0.802$\pm$0.010 \\
\enddata
\tablecomments{Asymptotic fluxes $F_{\nu}$ in the IRAC and MIPS bands. All values are in Jy, and are corrected for foreground galactic extinction. The quoted uncertainties are just the statitistical errors resulting from the fit applied to the growth curve, but do not include zero-point errors (see Section~\ref{asmag} for details).}
\end{deluxetable}

\clearpage
\begin{deluxetable}{rrrrrrrrrrrrrrrrrrrrrrr}
\tabletypesize{\scriptsize}
\setlength{\tabcolsep}{0.04in}
\tablecolumns{23}
\tablecaption{Non-parametrical morphological estimators.\label{allmorpho_tab}}
\tablewidth{0pt}
\rotate
\tablehead{
\colhead{Galaxy} & \colhead{} & \colhead{FUV} & \colhead{NUV} & \colhead{$B$} & \colhead{$V$} & \colhead{$R$} & \colhead{$I$} & \colhead{$u$} & \colhead{$g$} & \colhead{$r$} & \colhead{$i$} & \colhead{$z$} & \colhead{$J$} & \colhead{$H$} & \colhead{$K_S$} & \colhead{3.6\,$\micron$} & \colhead{4.5\,$\micron$} & \colhead{5.8\,$\micron$} & \colhead{8.0\,$\micron$} & \colhead{24\,$\micron$} & \colhead{70\,$\micron$} & \colhead{160\,$\micron$}}
\startdata
NGC~0925 & $C_{42}$ &  2.55 &  2.63 &  2.84 &  2.90 &  2.99 &  3.14 & \nodata & \nodata & \nodata & \nodata & \nodata &  3.07 &  3.02 &  3.05 &  3.27 &  3.25 &  3.40 &  3.51 &  3.14 &  3.01 &  2.52 \\
SAB(s)d & $A$ & 0.404 & 0.350 & 0.176 & 0.109 & 0.133 & 0.095 & \nodata & \nodata & \nodata & \nodata & \nodata & 0.015 & 0.010 & 0.000 & 0.149 & 0.123 & 0.134 & 0.260 & 0.309 & 0.224\dag & 0.205\dag \\
$T=7$ & $G$ & 0.648 & 0.619 & 0.551 & 0.516 & 0.550 & 0.530 & \nodata & \nodata & \nodata & \nodata & \nodata & 0.516 & 0.482 & 0.467 & 0.600 & 0.599 & 0.616 & 0.676 & 0.692 & 0.615\dag & 0.509\dag \\
& $\overline{M}_{20}$ & $-0.90$ & $-0.95$ & $-1.21$ & $-1.41$ & $-1.42$ & $-1.64$ & \nodata & \nodata & \nodata & \nodata & \nodata & $-1.75$ & $-1.75$ & $-1.58$ & $-1.56$ & $-1.41$ & $-1.27$ & $-1.28$ & $-0.85$ & $-1.15$\dag & $-1.62$\dag \\
\hline
NGC~2841 & $C_{42}$ &  2.28 &  2.36 &  3.77 &  3.93 &  3.93 &  3.89 &  3.33 &  3.84 &  3.92 &  3.91 &  3.96 &  3.84 &  3.84 &  3.84 &  3.95 &  3.83 &  3.12 &  2.00 &  1.82 &  1.90 &  2.15 \\
SA(r)b & $A$ & 0.289 & 0.213 & 0.107 & 0.082 & 0.065 & 0.052 & 0.084 & 0.098 & 0.070 & 0.054 & 0.031 & 0.009 & 0.008 & 0.000 & 0.027 & 0.018 & 0.022 & 0.086 & 0.084 & 0.041\dag & 0.081\dag \\
$T=3$ & $G$ & 0.586 & 0.586 & 0.595 & 0.606 & 0.610 & 0.605 & 0.566 & 0.611 & 0.626 & 0.629 & 0.647 & 0.666 & 0.662 & 0.667 & 0.648 & 0.645 & 0.633 & 0.601 & 0.612 & 0.610\dag & 0.444\dag \\
& $\overline{M}_{20}$ & $-1.02$ & $-1.15$ & $-2.42$ & $-2.47$ & $-2.45$ & $-2.51$ & $-2.16$ & $-2.48$ & $-2.53$ & $-2.55$ & $-2.61$ & $-2.58$ & $-2.55$ & $-2.57$ & $-2.56$ & $-2.54$ & $-2.03$ & $-1.09$ & $-0.99$ & $-1.02$\dag & $-1.04$\dag \\
\enddata
\tablecomments{Concentration index ($C_{42}$), asymmetry ($A$), Gini coeffient ($G$) and normalized second-order moment of the brightest 20\% to the total galaxy flux ($\overline{M}_{20}$) at different wavelengths. When the inner radius $r_{20}$ needed to compute $C_{42}$ is smaller than the innermost point in our profiles (or the PSF size, in the case of the 70\,$\micron$ and 160\,$\micron$ bands), the quoted values are just lower limits for the actual concentration indices. Values of $A$, $G$ and $\overline{M}_{20}$ marked with a dagger (\dag) should be taken with caution, since at the particular distance of each galaxy, the FWHM of the PSF of the affected bands does not allow resolving structures smaller than 0.5\,kpc.}
\end{deluxetable}

\clearpage
\begin{deluxetable}{rcccc}
\tabletypesize{\scriptsize}
\tablecolumns{5}
\tablecaption{Optical recalibration factors.\label{recalib_tab}}
\tablewidth{0pt}
\tablehead{
\colhead{Galaxy} & \colhead{$C_B$} & \colhead{$C_V$} & \colhead{$C_R$} & \colhead{$C_I$}\\
\colhead{(1)} & \colhead{(2)} & \colhead{(3)} & \colhead{(4)} & \colhead{(5)}}
\startdata
        NGC~0024 & 0.947 $\pm$ 0.008 & 1.041 $\pm$ 0.013 & 1.104 & 2.335 $\pm$ 0.011 \\
        NGC~0584 & 0.812 $\pm$ 0.006 & 0.951 $\pm$ 0.002 & 1.150 $\pm$ 0.003 & 1.604 $\pm$ 0.019 \\
        NGC~0925 & 0.844 $\pm$ 0.101 & 0.902 $\pm$ 0.125 & \nodata & 0.871 $\pm$ 0.003 \\
        NGC~1097 & 1.031 $\pm$ 0.019 & 0.970 $\pm$ 0.013 & 1.312 $\pm$ 0.017 & 1.663 $\pm$ 0.003 \\
        NGC~1291 & 1.034 $\pm$ 0.013 & 0.943 $\pm$ 0.008 & 1.370 & 1.738 \\
        NGC~1316 & 1.026 $\pm$ 0.009 & 0.932 $\pm$ 0.006 & 1.247 $\pm$ 0.028 & 1.571 $\pm$ 0.010 \\
        NGC~1377 & 2.339 $\pm$ 0.154 & 1.560 & 2.152 & 1.825 \\
        NGC~1404 & 1.082 $\pm$ 0.012 & 0.978 $\pm$ 0.008 & 1.272 $\pm$ 0.016 & 1.820 $\pm$ 0.013 \\
\enddata
\tablecomments{Recalibration factors for the SINGS optical data, derived from the compilation of aperture photometry of Prugniel \& Heraudeau (1998) (see Appendix~\ref{recalib}). The fluxes obtained directly from the SINGS images should be multiplied by these factors in order to get values closer to the actual ones. The quoted uncertainties result from the classical statistical formulas for linear fits, and do not include the intrinsic calibration errors of the data from the literature. We estimate that the final recalibrated data carry a global zero-point uncertainty of 10-15\%. Missing uncertainties correspond to cases in which there were not enough avalibale data-points to compute the statistical error. Some recalibration factors could not be computed when the aperture photometry catalog did not quote any measured flux for that particular galaxy and band. \dag The $R$-band image of NGC~3621 was delivered uncalibrated. This recalibration factor was derived after setting $\mathrm{\texttt{PHOTFLAM}}=10^{-7}$ in the FITS header.}
\end{deluxetable}

\clearpage

\clearpage

\clearpage

\clearpage

\clearpage

\clearpage

\clearpage

\clearpage

\clearpage

\clearpage
\begin{figure}
\resizebox{!}{0.8\vsize}{\includegraphics{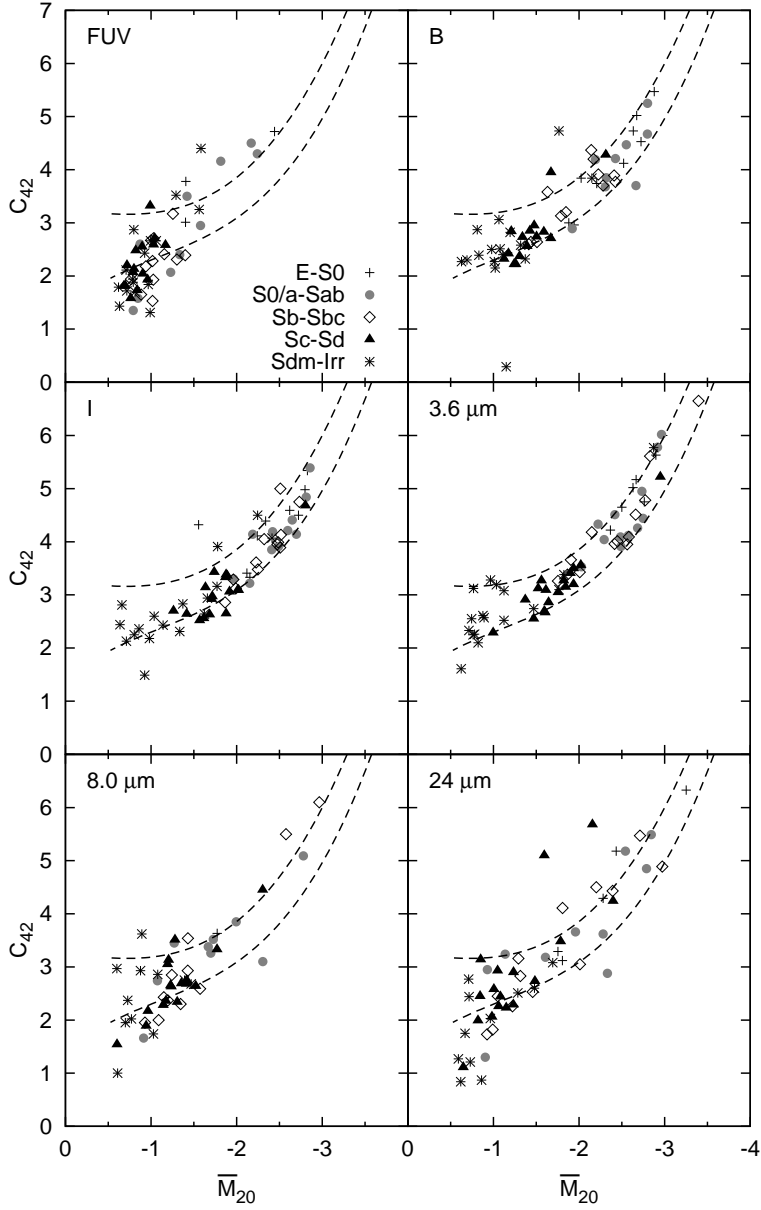}}
\caption{Concentration index of the SINGS galaxies as a function of
the normalized second-order moment of the brightest 20\% of the
emission. Trends are shown at selected bands, using different
symbols to sort out galaxies into different Hubble types. The dashed
lines are third-order polynomials that fit the upper and lower
envelopes of the data at 3.6\,$\micron$, and are replicated in all
panels to facilitate the visual comparison of the trends at different
wavelengths.\label{conc_vs_m20}}
\end{figure}

\clearpage
\begin{figure}
\resizebox{!}{0.8\vsize}{\includegraphics{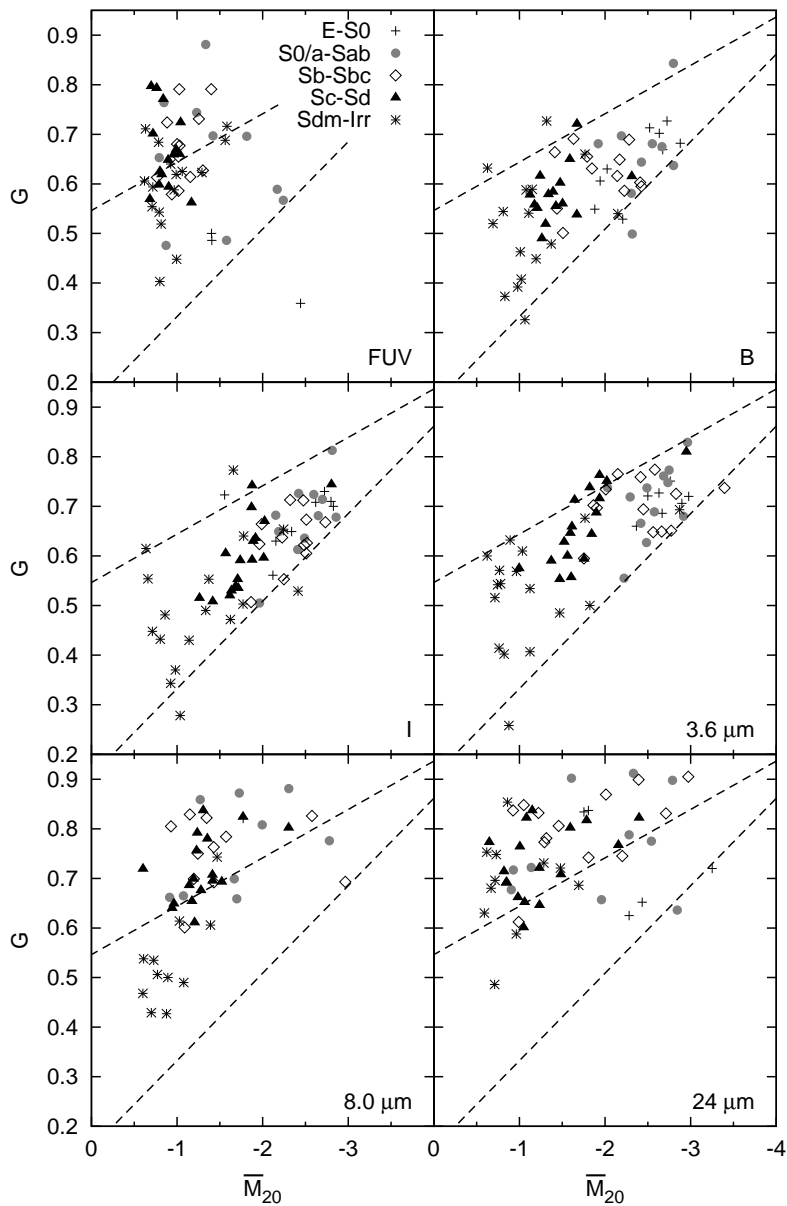}}
\caption{Gini coefficient of the SINGS galaxies as a function of the
normalized second-order moment of the brightest 20\% of the
emission. Each panel shows a different band, and Hubble types are
coded with different symbols. The upper and lower boundaries of the
data-cloud at 3.6\,$\micron$ are fitted with two straight lines, which
are replicated in all panels for the ease of
comparison.\label{gini_vs_m20}}
\end{figure}

\clearpage
\begin{figure}
\resizebox{!}{0.8\vsize}{\includegraphics{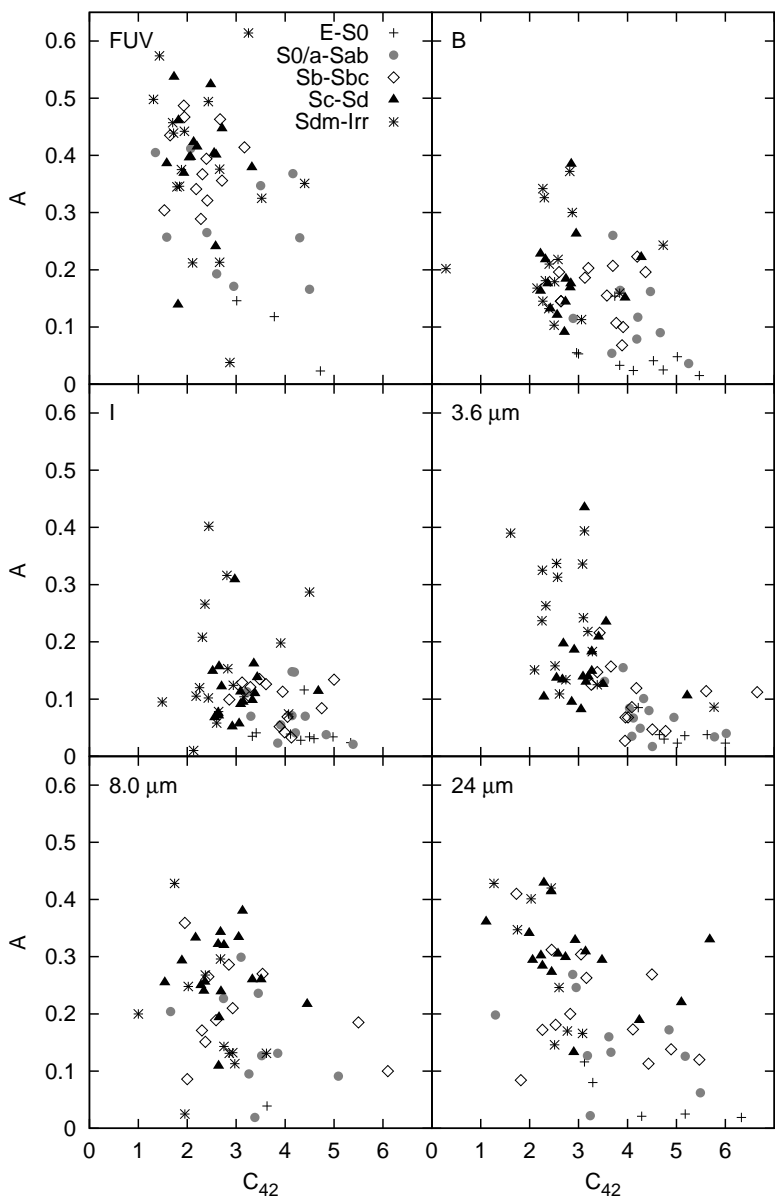}}
\caption{Asymmetry of the SINGS galaxies as a function of their
concentration indices at several wavelengths. Different symbols are
used to sort out galaxies into different Hubble types. Note that, as
explained in Section~\ref{asymmetry}, the systematically large
asymmetries displayed by Sdm and irregular galaxies at 3.6\,$\micron$
is most probably due to the contribution of foreground and background
sources, since galaxies of these Hubble types usually have low surface
brightness in the near-IR.\label{conc_vs_asym}}
\end{figure}

\clearpage
\begin{figure}
\resizebox{!}{0.8\vsize}{\includegraphics{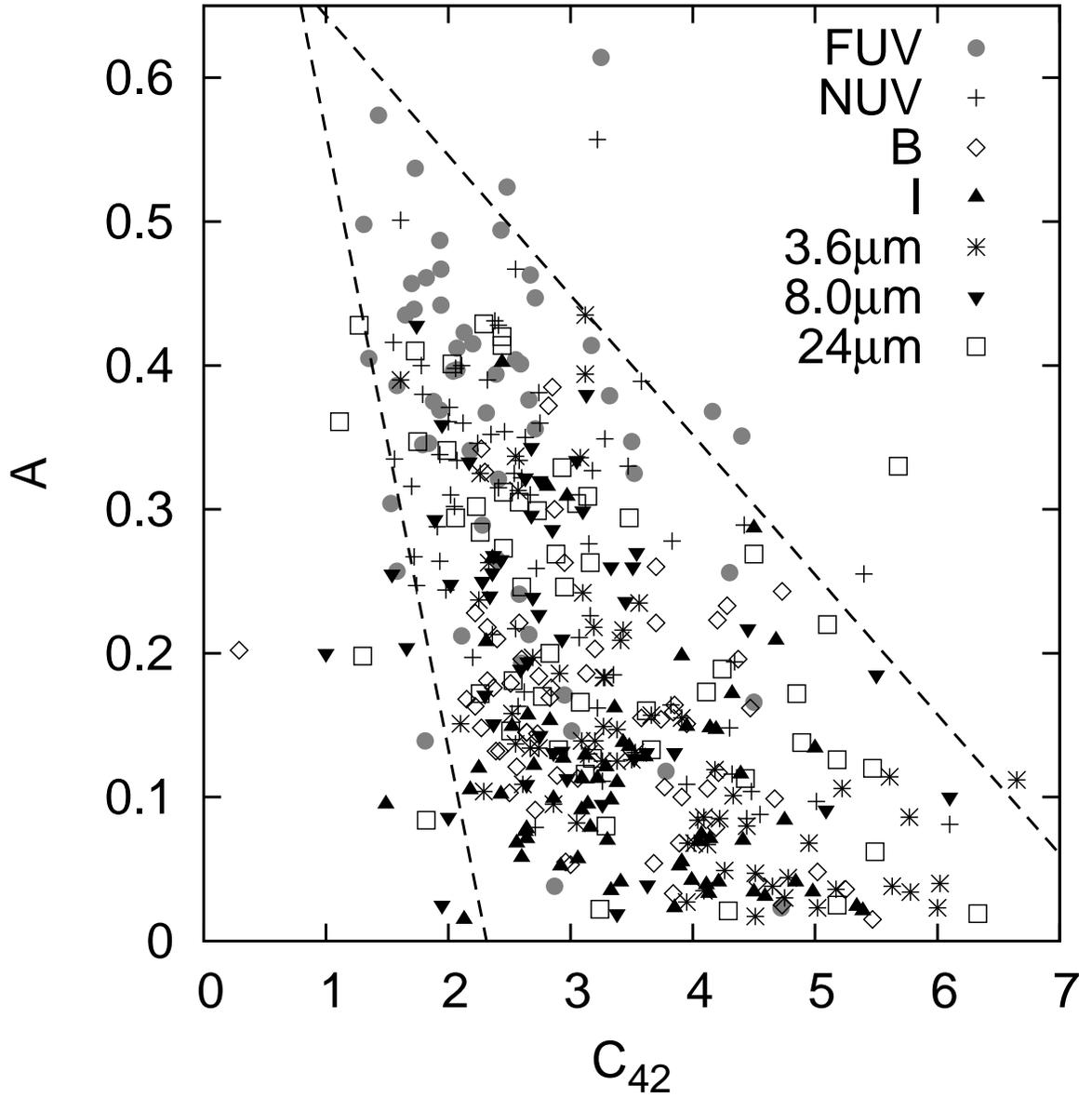}}
\caption{Asymmetry as a function of the concentration index for all
galaxies and several bands displayed at the same time. The upper and
lower limits were obtained by fitting the boundaries of the data-point
distribution in all the quoted bands
simultaneously.\label{conc_vs_asym_all}}
\end{figure}

\clearpage
\begin{figure}
\resizebox{1\hsize}{!}{\includegraphics{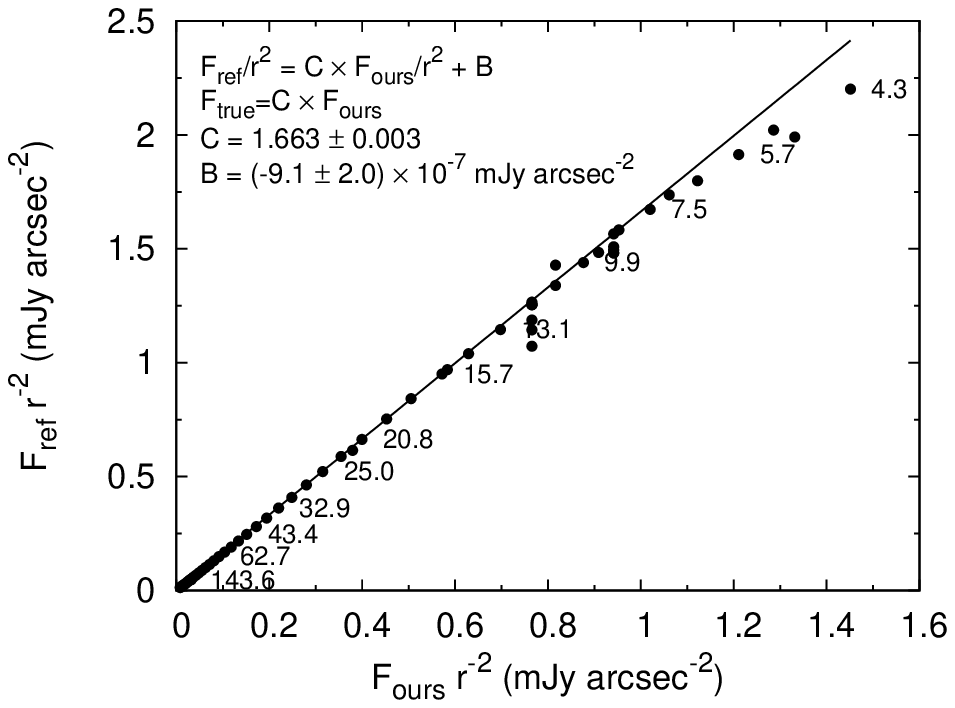}}
\caption{Sample recalibration plot for the I-band image of
NGC~1097. We compare the flux enclosed inside circular apertures of
radius $r$ measured on our images with published aperture photometry
(see Appendix~\ref{recalib}). The small numbers next to each point
show the radius in arcseconds of the corresponding aperture (not all
of them are shown for clarity). Very small apertures do not always
follow the linear trend, likely due to differences in the PSF and
centering errors, and are thus excluded from the fitting.
\label{ngc1097_recalib}}
\end{figure}


\begin{thebibliography}{}
\bibitem[]{} Abraham, R. G., Tanvir, N. R., Santiago, B. X., Ellis, R. S., Glazebrook, K., \& van den Bergh, S. 1996a, MNRAS, 279, 47
\bibitem[]{} Abraham, R. G., van den Bergh, S., Glazebrook, K., Ellis, R. S., Santiago, B. X., Surma, P., \& Griffiths, R. E. 1996b, ApJS, 107, 1
\bibitem[]{} Abraham, R. G.; van den Bergh, S., \& Nair, P. 2003, ApJ, 588, 218
\bibitem[]{} Adelman-McCarthy, J. K., et al. 2008, ApJS, 175, 297
\bibitem[]{} Baggett, W. E., Baggett, S. M., \& Anderson, K. S. J. 1998, AJ, 116, 1626
\bibitem[]{} Bakos, J., Trujillo, I., \& Pohlen, M. 2008, ApJ, 683, 103
\bibitem[]{} Bendo, G. J., et al. 2006, ApJ, 645, 134
\bibitem[]{} Bendo, G. J., et al. 2007, MNRAS, 380, 1313
\bibitem[]{} Bershady, M. A., Jangren, A., \& Conselice, C. J. 2000, AJ, 119, 2645
\bibitem[]{} Bertin, E., \& Arnouts, S. 1996, A\&AS, 117, 393
\bibitem[]{} Boissier, S., Boselli, A., Buat, V., Donas, J., \& Milliard, B. 2004, A\&A, 424, 465
\bibitem[]{} Boissier, S., \& Prantzos, N. 2000, MNRAS, 312, 398
\bibitem[]{} Boissier, S., et al. 2005, ApJ, 619, 83
\bibitem[]{} Boissier, S., et al. 2007, ApJS, 173, 524
\bibitem[]{} Boselli, A., Tuffs, R. J., Gavazzi, G., Hippelein, H., \& Pierini, D. 1997, A\&AS, 121, 507
\bibitem[]{} Buat, V., et al. 2005, ApJ, 619, 51
\bibitem[]{} Burgarella, D., Buat, V., Donas, J., Milliard, B., \& Chapelon, S. 2001, A\&A, 369, 421
\bibitem[]{} Bush, S. J., Cox, T. J., Hernquist, L., Thilker, D., \& Younger, J. D. 2008, ApJ, 683, 13
\bibitem[]{} Cair\'{o}s, L. M., Caon, N., Vílchez, J. M., Gonz\'{a}lez-P\'{e}rez, J. N., \& Mu\~{n}oz-Tu\~{n}\'{o}n, C. 2001, ApJS, 136, 393
\bibitem[]{} Calzetti, D., Kinney, A. L., \& Storchi-Bergmann, T. 1994, ApJ, 429, 582
\bibitem[]{} Cardiel, N. 2009, MNRAS, 396, 680
\bibitem[]{} Cohen, M., Wheaton, Wm. A., \& Megeath, S. T. 2003, AJ, 126, 1090
\bibitem[]{} Conselice, C. J., Bershady, M. A., \& Jangren, A. 2000, ApJ 529, 886
\bibitem[]{} Cortese, L., Boselli, A., Franzetti, P., Decarli, R., Gavazzi, G., Boissier, S., \& Buat, V. 2008, MNRAS, 386, 1157
\bibitem[]{} Cutri, R. M., et al. 2003, Explanatory Supplement to the 2MASS All Sky Data Release and Extended Mission Products, http://www.ipac.caltech.edu/2mass/releases/allsky/doc/explsup.html
\bibitem[]{} Dale, D. A., et al. 2007, ApJ, 655, 863
\bibitem[]{} de Jong, R.S. 1996, A\&A, 313, 377
\bibitem[]{} de Vaucouleurs, G. 1958, ApJ, 128, 465
\bibitem[]{} de Vaucouleurs, G., 1977, Evolution of galaxies and stellar populations, ed. R. B. Larson, \& B. M., Tynsley, Yale Univ. Obs., New Haven, 43
\bibitem[]{} de Vaucouleurs, G., de Vaucouleurs, A., Corwin, H.G., Buta, R.J., Paturel, G., \& Fouqu\'{e}, P. 1991, Third Reference Catalogue of Bright Galaxies (RC3) (Springer-Verlag)
\bibitem[]{} Dellenbusch, K. E., Gallagher, J. S., III, \& Knezek, P. M. 2007, ApJ, 655, 29
\bibitem[]{} Draine, B. T., \& Li, A. 2007, ApJ, 657, 810
\bibitem[]{} Draine, B. T., et al. 2007, ApJ, 663, 866 
\bibitem[]{} Drozdovsky, I. O., \& Karachentsev, I. D. 2000, A\&AS, 142, 425
\bibitem[]{} Engelbracht, C. W., Gordon, K. D., Rieke, G. H., Werner, M. W., Dale, D. A., \& Latter, W. B. 2005, ApJ, 628, 29
\bibitem[]{} Engelbracht, C. W., Rieke, G. H., Gordon, K. D., Smith, J. D. T., Werner, M. W., Moustakas, J., Willmer, C. N. A., \& Vanzi, L. 2008, ApJ, 678, 804
\bibitem[]{} Engelbracht, C. W., et al. 2007, PASP, 119, 994
\bibitem[]{} Erwin, P., Pohlen, M., \& Beckman, J. E. 2008, AJ, 135, 20
\bibitem[]{} Fall, S. M., \& Efstathiou, G. 1980, MNRAS, 193, 189
\bibitem[]{} Fazio, G. G., et al. 2004, ApJS, 154, 10
\bibitem[]{} Ferguson, A. M. N., \& Clarke, C. J. 2001, MNRAS, 325, 781
\bibitem[]{} Freeman, K. C. 1970, ApJ, 160, 811
\bibitem[]{} Fukugita, M., Shimasaku, K., \& Ichikawa, T. 1995, PASP, 107, 945
\bibitem[]{} Gil de Paz, A., \& Madore, B. F. 2005, ApJS, 156, 345
\bibitem[]{} Gil de Paz, A., et al. 2005, ApJ, 627, 29
\bibitem[]{} Gil de Paz, A., et al. 2007, ApJS, 173, 185
\bibitem[]{} Gini, C. 1912, reprinted in Memorie di Metodologia Statistica, ed. E. Pizetti \& T. Salvemini (1955; Rome: Libreria Eredi Virgilio Veschi)
\bibitem[]{} Gordon, K. D., Clayton, G. C., Witt, A. N., \& Misselt, K. A. 2000, ApJ, 533, 236
\bibitem[]{} Gordon, K. D., et al. 2007, PASP, 119, 1019
\bibitem[]{} Governato, F., Willman, B., Mayer, L., Brooks, A., Stinson, G., Valenzuela, O., Wadsley, J., \& Quinn, T. 2007, MNRAS, 374, 1479
\bibitem[]{} Graham, A. W., Driver, S. P., Petrosian, V., Conselice, C. J., Bershady, M. A., Crawford, S. M., \& Goto, T. 2005, AJ, 130, 1535
\bibitem[]{} Heyer, M. H., Corbelli, E., Schneider, S. E., \& Young, J. S. 2004, ApJ, 602, 723
\bibitem[]{} Jarrett, T. H., Chester, T., Cutri, R., Schneider, S. E., \& Huchra, J. P. 2003, AJ, 125, 525
\bibitem[]{} Kennicutt, R. C., Jr., et al. 2003, PASP, 115, 928
\bibitem[]{} Kent, S. M. 1985, ApJS, 59, 115
\bibitem[]{} Kormendy, J. 1977, ApJ, 217, 406
\bibitem[]{} Kuchinski, L. E., Madore, B. F., Freedman, W. L., \& Trewhella, M. 2001, AJ, 122, 729
\bibitem[]{} Kuchinski, L. E., et al. 2000, ApJS, 131, 441
\bibitem[]{} Lauger, S., Burgarella, D., \& Buat, V. 2005, A\&A, 434, 77
\bibitem[]{} Li, A., \& Draine, B. T. 2001, ApJ, 554, 778
\bibitem[]{} L\'{o}pez-Sanjuan, C. et al. 2009, ApJ, 694, 643
\bibitem[]{} Lotz, J. M., Primack, J., \& Madau, P. 2004, AJ, 128, 163
\bibitem[]{} MacArthur, L. A., Courteau, S., Bell, E., \& Holtzman, J. A. 2004, ApJS, 152, 175
\bibitem[]{} Marcum, P. M., et al. 2001, ApJS, 132, 129
\bibitem[]{} Martin. D., et al. 2005, ApJ, 619 1
\bibitem[]{} Mu\~{n}oz-Mateos, J. C., et al. 2007, ApJ, 658, 1006
\bibitem[]{} Mu\~{n}oz-Mateos, J. C., et al. 2009, ApJ, 701, 1965 (Paper II)
\bibitem[]{} Ochsenbein, F., Bauer, P., \& Marcout, J. 2000, A\&AS, 143, 23
\bibitem[]{} O'Connell, R. W. 1999, ARA\&A, 37, 603
\bibitem[]{} Ohl, R.G., et al. 1998, ApJ, 505, 11
\bibitem[]{} Oke, J. B. 1974, ApJS, 27, 21
\bibitem[]{} Peimbert, M., \& Torres-Peimbert, S. 1981, ApJ, 245, 845
\bibitem[]{} Petrosian, V. 1976, ApJ, 209, 1
\bibitem[]{} Pohlen, M., \& Trujillo, I. 2006, A\&A, 454, 759
\bibitem[]{} Pohlen, M., et al. 2008, in ASP Conf. Ser. 396, Formation and Evolution of Galaxy Disks, ed. Funes, J. G., \& Corsini, E. M. (San Francisco, CA: ASP), 183
\bibitem[]{} Popescu, C. C., Misiriotis, A., Kylafis, N. D., Tuffs, R. J., \& Fischera, J. 2000, A\&A, 362, 138
\bibitem[]{} Prugniel, Ph., \& Heraudeau, Ph. 1998, A\&AS, 128, 299
\bibitem[]{} Reach, W. T., et al. 2005, PASP, 117, 978
\bibitem[]{} Rieke, G. H., et al. 2004, ApJS, 154, 25
\bibitem[]{} Robin, A. C., et al. 2007, ApJS, 172, 545
\bibitem[]{} R\u{o}skar, R., Debattista, V. P., Stinson, G. S., Quinn, T. R., Kaufmann, T., \& Wadsley, J. 2008, ApJ, 675, 65
\bibitem[]{} Scarlata, S., et al. 2007, ApJS, 172, 406
\bibitem[]{} Schade, D., Lilly, S. J., Crampton, D., Hammer, F., Le Fevre, O., \& Tresse, L. 1995, ApJ, 451, 1
\bibitem[]{} Schlegel, D. J., Finkbeiner, D. P., \& Davis, M. 1998, ApJ, 500, 525
\bibitem[]{} S\'ersic, J. L., 1968, Atlas de Galaxias Australes (C\'ordoba, Argentina: Observatorio Astron\'omico)
\bibitem[]{} Stansberry, J. A., et al. 2007, PASP, 119, 1038
\bibitem[]{} Taylor-Mager, V. A., Conselice, C. J., Windhorst, R. A., \& Jansen, R. A. 2007, ApJ, 659, 162
\bibitem[]{} Thilker, D. A., et al. 2005, ApJ, 619, 79
\bibitem[]{} Thilker, D. A., et al. 2007, ApJS, 173, 538
\bibitem[]{} van der Kruit, P. C. 1979, A\&AS, 38, 15
\bibitem[]{} Walter, F., Brinks, E., de Blok, W. J. G., Bigiel, F., Kennicutt, R. C., Jr., Thornley, M. D., \& Leroy, A. K. 2008, AJ, 136, 2563
\bibitem[]{} Werner, M. W., et al. 2004, ApJS, 154, 1
\bibitem[]{} Witt, A. N., \& Gordon, K. D. 2000, ApJ, 528, 799
\bibitem[]{} Wong, T., \& Blitz, L. 2002, ApJ, 569, 157
\bibitem[]{} Xilouris, E. M., Byun, Y. L., Kylafis, N. D., Paleologou, E. V., Papamastorakis, \& J. 1999, A\&A, 344, 868
\bibitem[]{} York, D., et al. 2000, AJ, 120, 1579
\bibitem[]{} Yoshii, Y., \& Sommer-Larsen, J. 1989, MNRAS, 236, 779
\end{thebibliography}
\end{document}